\documentclass[a4paper,11pt]{article}
\usepackage[T1]{fontenc}
\usepackage{textcomp}
\usepackage{jheppub}
\usepackage{amsmath}
\usepackage{graphicx}
\usepackage{appendix}
\usepackage{titlesec}
\usepackage{subcaption}
\usepackage[normalem]{ulem}
\usepackage{xcolor}
\usepackage{enumitem}
\usepackage{booktabs}
\usepackage{siunitx}
\usepackage{pifont}
\usepackage{needspace}
\newcommand{\cmark}{\ding{51}}%
\newcommand{\xmark}{\ding{55}}%
\newcommand{\ii}{\mathrm{i}}

\renewcommand{\thefootnote}{\fnsymbol{footnote}}

\definecolor{mygreen}{RGB}{34,139,34}

\usepackage{lineno}
\title{Interpreting the 95 GeV di-photon and di-tau Excesses in the Georgi-Machacek Model}
\author[1,2,3]{Qin Chang,}
\author[1,2]{Xiaokang Du,\footnote{Corresponding author.}}
\author[4]{Pengxuan Zhu}

\affiliation[1]{Institute of Physics, Henan Academy of Sciences, Zhengzhou 450046, P. R. China}
\affiliation[2]{Centre for Theoretical Physics, Henan Normal University, Xinxiang 453007, P. R. China}
\affiliation[3] {Center for High Energy Physics, Henan Academy of Sciences, Zhengzhou 455004, P. R. China}
\affiliation[4] {ARC Centre of Excellence for Dark Matter Particle Physics \& CSSM,
Department of Physics, Adelaide University, Adelaide, SA 5005, Australia}

\emailAdd{xkdu@hnas.ac.cn}
\emailAdd{changqin@htu.edu.cn}
\emailAdd{pengxuan.zhu@adelaide.edu.au}

\abstract{We revisit the 95~GeV $\gamma\gamma$ and $\tau\tau$ excesses in the Georgi--Machacek model using a combined fit to a single light $CP$-even custodial singlet $H$. Using the one-loop renormalization group-improved effective potential and positive-definiteness conditions for vacuum stability, together with perturbative unitarity, electroweak precision tests, $B$-physics, and Higgs data, we identify a narrow but viable parameter region. In the allowed region $\mu_{\gamma\gamma}$ approaches the experimental central value, while $\mu_{\tau\tau}$ reaches only $\sim0.5$ (about $1.4\sigma$ below the CMS central value) and remains within the $2\sigma$ interval---a good description of the di-photon excess but only a partial accommodation of the di-tau excess. The scenario predicts characteristic patterns in the singlet mixing and triplet vacuum expectation values and is highly testable at the HL-LHC and future lepton colliders via precision $\kappa_V$ measurements and direct exotic searches.}

\begin{document}
\maketitle

\newpage

\section{Introduction}
\label{sec-1}
\renewcommand{\thefootnote}{\textsuperscript{\arabic{footnote}}}

The discovery of the 125 GeV Higgs boson by the ATLAS \cite{ATLAS:2012ae} and CMS \cite{CMS:2012zhx} Collaborations in 2012 confirms the Standard Model (SM) framework of electroweak symmetry breaking (EWSB). Nevertheless, theoretical motivations and the sizable uncertainties in current Higgs data fits still leave ample room for an extended Higgs sector \cite{ATLAS:2019nkf}. Consequently, the search for additional scalar fields remains a crucial objective.

\par 
The CMS and ATLAS Collaborations have each reported a di-photon excess around 95.4 GeV \cite{CMS:2024yhz, CMS:2018cyk, ATLAS:2024bjr}. A phenomenological combination of these Run-2 di-photon results in \cite{Biekotter:2023oen} finds a local significance of $3.1\sigma$ and a combined signal strength $\mu_{\gamma\gamma}^{\rm ATLAS+CMS} = 0.24^{+0.09}_{-0.08}$.
CMS has also observed another scalar resonance at 100 GeV with a local significance of $3.1\sigma$ \cite{CMS:2022goy}.
Because the mass resolution in the di-tau channel is poorer than in the di-photon channel, this excess could be compatible with the same scalar as the di-photon excess. At 95 GeV, the local significance in the di-tau channel is 2.6$\sigma$, with a signal strength of $\mu_{\tau\tau}^{\rm CMS} = 1.2 \pm 0.5$.
Additionally, data from the Large Electron-Positron Collider (LEP) show an excess at 98 GeV in the $b\bar{b}$ channel, with a local significance of 2.3$\sigma$ and a signal strength of $\mu_{b\bar{b}}^{\rm LEP} = 0.117 \pm 0.057$ \cite{LEP:2003ing}.
\par Although each excess is individually modest in significance, their persistence motivates systematic exploration within well-defined new physics frameworks. Since the three excesses cluster in a similar mass region and exhibit signal strengths comparable to those expected from a SM-like Higgs boson, it is natural to view them as potential hints of an extended Higgs sector in the context of EWSB. It is therefore important to examine whether they can be consistently interpreted as manifestations of a light scalar particle.
There is broad theoretical interest in simultaneously explaining the di-photon, di-tau, and $b\bar{b}$ excesses within a consistent framework.~\cite{Biekotter:2023oen, Heinemeyer:2018jcd, Heinemeyer:2018wzl}.
These include the scalar extensions~\cite{Kundu:2019nqo, Bhattacharya:2023lmu}, dilation model~\cite{Liu:2018ryo, Wang:2024bkg}, extra-${\rm U(1)}$ symmetry~\cite{Liu:2018xsw, Aguilar-Saavedra:2020wrj, Aguilar-Saavedra:2023tql, Ge:2024rdr, Baek:2024cco}, left-right model \cite{Dev:2023kzu}, triplet Higgs \cite{Ashanujjaman:2023etj}, Scotogenic model~\cite{Escribano:2023hxj, Borah:2023hqw, Ahriche:2023hho}, Higgs radion model~\cite{Richard:2017kot, Sachdeva:2019hvk} and other mechanisms~\cite{Abbas:2024jut, Li:2025tkm, YaserAyazi:2024hpj}.
Discussions have also been carried out in various versions of the two-Higgs-doublet model (2HDM)~\cite{Fox:2017uwr, Haisch:2017gql, Benbrik:2022azi, Benbrik:2022dja,  Azevedo:2023zkg, Belyaev:2023xnv, Dutta:2023cig, Coloretti:2023yyq, Benbrik:2024ptw, Khanna:2024bah, BrahimAit-Ouazghour:2024img, Banik:2024ugs, Benbrik:2025hol, Khanna:2025cwq}, and in the singlet-extended 2HDM~\cite{Arcadi:2023smv, Gao:2024qag, Biekotter:2019kde, Biekotter:2019mib, Biekotter:2019drs, Biekotter:2020ahz, Biekotter:2020cjs, Biekotter:2021qbc, Heinemeyer:2021msz, Biekotter:2022jyr, Biekotter:2023qbe, Xu:2025vmy}.
In a supersymmetric framework, the minimal supersymmetric standard model (MSSM) cannot provide a common explanation~\cite{Bechtle:2016kui}.
By contrast, a well-motivated supersymmetric explanation of the observed excesses is provided by the next-to-minimal supersymmetric standard model (NMSSM), in which a mostly singlet-like Higgs boson can naturally appear around 95~GeV~\cite{Cao:2016uwt, Choi:2019yrv, Cao:2019ofo, Li:2022etb, Ellwanger:2023zjc, Cao:2023gkc, Li:2023kbf, Cao:2024axg, Ellwanger:2024txc, Ellwanger:2024vvs, Lian:2024smg}.
Another concrete example is offered by the $\mu\nu$SSM, where the $CP$-even right-handed sneutrino may play the role of a light scalar explaining the excesses~\cite{Liu:2024cbr, Biekotter:2019gtq}.
Beyond these cases, further supersymmetric scenarios, such as the minimal $R$-symmetry supersymmetric standard model (MRSSM)~\cite{Kalinowski:2024uxe, Kotlarski:2024eyw}, have also been explored as a possible solution to accommodate a light scalar state.
Finally, several works have approached from the collider perspective~\cite{Coloretti:2023wng, Biekotter:2023jld, Banik:2023vxa, Mondal:2024obd, Dong:2025orv, Janot:2024ryq, Belanger:2024wca}.

\par The Georgi-Machacek (GM) model~\cite{Georgi:1985nv, Chanowitz:1985ug, Gunion:1989ci, Gunion:1990dt, Gunion:1989we, Haber:1999zh, Godfrey:2010qb, Logan:2010en, Low:2012rj, Killick:2013mya, Efrati:2014uta, Degrande:2015xnm} extends the SM by introducing a complex ${\rm SU(2)}_L$ triplet scalar $\chi$ with hypercharge $+1$ and a real ${\rm SU(2)}_L$ triplet scalar $\xi$ with hypercharge $0$.
The field content is arranged such that custodial symmetry is preserved at tree level after EWSB, provided the triplet vacuum expectation values (VEVs) are aligned, $v_\chi = v_\xi \equiv v_\Delta$.
This alignment permits $v_\Delta$ values up to several tens~GeV, significantly larger than in generic triplet extensions, while maintaining $\rho_{\rm tree}=1$.
The model predicts nine additional scalar states, including a characteristic doubly charged Higgs boson, resulting in a rich phenomenology and distinctive collider signatures~\cite{Chiang:2012cn, Kanemura:2013mc, Englert:2013zpa, Englert:2013wga, Godunov:2014waa, Chiang:2015kka, Chiang:2015amq, Yu:2016skv, Cao:2016rht, Chang:2017niy, Zhang:2017och, Li:2017daq, Sun:2017mue, Azuelos:2017dqw, Cao:2018xwy, Yang:2018pzt, Ghosh:2019qie, Ismail:2020zoz, Zhu:2020bew, Wang:2022okq, Ghosh:2022wbe, deLima:2022yvn, Ahriche:2022aoj, Ghosh:2022bvz, Chakraborti:2023mya, Ghosh:2023izq, Bhattacharya:2024gyh, Lu:2024ade, Ghosh:2025htt, Ashanujjaman:2025puu, Ghosh:2025hyy, Chiang:2025lab, Chiang:2018cgb}.
Notably, the GM model allows the couplings of the SM-like Higgs boson to weak gauge bosons to exceed their SM values~\cite{Falkowski:2012vh, Chang:2012gn, Chiang:2013rua}.

\par The extended scalar sector improves vacuum stability and can induce strong first-order electroweak phase transitions (EWPT), potentially generating gravitational wave signals accessible to future experiments~\cite{Chiang:2014hia, Chen:2022zsh, Zhou:2018zli, Bian:2019bsn, Lu:2025vif}.
If lepton number violation is present, triplet-lepton interactions can generate neutrino masses~\cite{Chen:2020ark, Giarnetti:2023dcr, Giarnetti:2023osf, Du:2022brr}.
The GM framework has also been employed to address the $W$-mass anomaly reported by CDF-II~\cite{Du:2022brr, Mondal:2022xdy, Chen:2022ocr, Ghosh:2022zqs}, especially when custodial symmetry breaking effects are included~\cite{Blasi:2017xmc, Keeshan:2018ypw, Song:2022jns}.
Furthermore, the light custodial singlet scalar of the GM model constitutes a natural candidate to interpret the excesses observed near 95~GeV~\cite{Ahriche:2023wkj, Chen:2023bqr, Du:2025eop, Ghosh:2025ebc, Mondal:2025tzi}.
An explanation for the excesses observed in the $\gamma\gamma$, $\tau\tau$, and $b\bar{b}$ channels requires a near-degenerate $CP$-even/$CP$-odd `twin-peak' structure~\cite{Ahriche:2023wkj}; with only a single light $CP$-even state, the $\tau\tau$ excess is in tension with the $\gamma\gamma$ and $b\bar b$ channels~\cite{Ahriche:2023wkj, Chen:2023bqr, Du:2025eop}.
Recent analyses, however, strongly disfavor the LEP $b\bar{b}$ excess as evidence for a new scalar~\cite{Janot:2024ryq}, motivating a focus on the LHC-driven $\gamma\gamma$ and $\tau\tau$ channels.

The positive-definiteness vacuum stability conditions for the GM model were proposed in~\cite{Du:2024mry}, which preclude the existence of deeper vacua in large field value regions. Moving beyond the conventional tree-level analysis, the study employs the one-loop renormalization group-improved (RG-improved) effective potential, along with a novel criterion for the positive definiteness of homogeneous polynomials in multiple variables—a refinement necessitated by loop-induced custodial symmetry breaking effects. Through numerical analysis, it is shown that the quartic coupling ranges permitted by the positive-definiteness requirements significantly deviate from those derived from tree-level bounded-from-below conditions. Notably, certain parameter regions previously excluded by tree-level criteria become viable when the one-loop RG-improved scalar potential is considered. Conversely, some regions of parameter space that satisfy tree-level bounded-from-below constraints at the electroweak scale are ultimately ruled out by the positive-definiteness requirements of the effective potential in large field value regimes.

The distinctive aim of this work is to quantify how the one-loop RG-improved positive-definiteness condition affects a single light $CP$-even scalar interpretation of the LHC $\gamma\gamma$ and $\tau\tau$ indications. Relative to Ref.~\cite{Chen:2023bqr} (tree-level BFB), the viable di-tau strength is enlarged from $\mathcal{O}(0.1)$ to $\sim0.5$; relative to Ref.~\cite{Du:2025eop}, the LEP $b\bar b$ excess is not used as a fitting input; and unlike Ref.~\cite{Ahriche:2023wkj}, we retain one light $CP$-even scalar rather than a twin-peak construction. The CMS di-tau central value remains inaccessible because of tension with the $125~\mathrm{GeV}$ Higgs couplings.

This work is organized as follows. In Sec.~\ref{sec-2}, we provide a brief review of the GM model. A short discussion of the di-photon and di-tau excess signals around 95~GeV is presented in Sec.~\ref{sec-3}. Theoretical and experimental constraints relevant to realistic studies within the GM model are detailed in Sec.~\ref{sec-4}. Numerical results are displayed in Sec.~\ref{sec-5}. Finally, we present our conclusions in Sec.~\ref{sec:sum}.

%%%%%%%%%%%%%%%%%%%%%%%%%%%%%%%%%%%%%%%%%%%%%%%%%%%%%%%%%%%%
\section{The Georgi-Machacek model}
\label{sec-2}
\par The GM model extends the SM scalar sector by introducing additional triplet fields. It contains the familiar ${\rm SU(2)}_L$ doublet $({\phi}^+,~{\phi}^0)$ with hypercharge $1/2$, a complex triplet $({\chi}^{++},~{\chi}^{+},~{\chi}^{0})$ with hypercharge $+1$, and a real triplet $({\xi}^{+},~{\xi}^{0},~-{\xi}^{+*})$ with zero hypercharge. This field structure enables custodial ${\rm SU(2)}_V$ symmetry at tree level and leads to distinctive scalar phenomenology.
\par To clarify custodial symmetry, the doublet and triplets are arranged in matrices $\Phi$ and $\Delta$, respectively:
\begin{equation}\label{eq:Higgs_matrices}
  \Phi=
  \begin{pmatrix}
    \phi^{0*} & \phi^+ \\
    \phi^- & \phi^0
  \end{pmatrix}, \quad
  \Delta=
  \begin{pmatrix}
    \chi^{0*} & \xi^+ & \chi^{++} \\
    \chi^- & \xi^0 & \chi^{+} \\
    \chi^{--} & \xi^- & \chi^{0}
  \end{pmatrix}.
\end{equation}

\par Then the most general scalar potential for the GM scalar sector, constructed to be invariant under a global ${\rm SU(2)}_L \times {\rm SU(2)}_R$ symmetry, hence also under its gauged subgroup ${\rm SU(2)}_L \times {\rm U(1)}_Y$, reads
\begin{equation}
  \label{eq:GMpot}
  \begin{split}
    V(\Phi,\Delta) =
    & \frac12 m_{\Phi}^2 {\rm tr} \left[ \Phi^\dagger \Phi \right]
    + \frac{1}{2} m_{\Delta}^2 {\rm tr}\left[ \Delta^\dagger \Delta \right] \\
    &  + \lambda_1 \left( {\rm tr} \left[ \Phi^\dagger \Phi \right] \right)^2
    + \lambda_2 \left( {\rm tr} \left[ \Delta^\dagger \Delta \right] \right)^2
    + \lambda_3 {\rm tr} \left[ \left( \Delta^\dagger \Delta \right)^2 \right] \\
    &  + \lambda_4 {\rm tr} \left[ \Phi^\dagger \Phi \right] {\rm tr} \left[ \Delta^\dagger \Delta \right]
    + \lambda_5 {\rm tr} \left[ \Phi^\dagger \frac{\sigma^a}{2} \Phi \frac{\sigma^b}{2} \right] {\rm tr} \left[ \Delta^\dagger T^a \Delta T^b \right] \\
    &   + \mu_1 {\rm tr} \left[ \Phi^\dagger \frac{\sigma^a}{2} \Phi \frac{\sigma^b}{2} \right] (P^\dagger \Delta P)_{ab}
    + \mu_2 {\rm tr} \left[ \Delta^\dagger T^a \Delta T^b \right] (P^\dagger \Delta P)_{ab},
  \end{split}
\end{equation}
Here, $\sigma^i$ are Pauli matrices, and $T^i$ are the ${\rm SU(2)}$ generators in the triplet representation:
\begin{equation}
  T^1 = \frac{1}{\sqrt{2}}
  \begin{pmatrix}
    0 & 1 & 0 \\
    1 & 0 & 1 \\
    0 & 1 & 0
  \end{pmatrix},\quad
  T^2 = \frac{1}{\sqrt{2}}
  \begin{pmatrix} 0 & -\ii & 0 \\ \ii & 0 & -\ii \\ 0 & \ii & 0
  \end{pmatrix}, \quad
  T^3 =
  \begin{pmatrix} 1 & 0 & 0 \\ 0 & 0 & 0 \\ 0 & 0 & -1
  \end{pmatrix}.
\end{equation}
The $P$ matrix rotates the triplet fields $\Delta$ to the Cartesian basis
\begin{align}
  P = & \frac{1}{\sqrt{2}}
  \begin{pmatrix} -1 & \ii & 0 \\ 0 & 0 & \sqrt{2} \\ 1 & \ii & 0
  \end{pmatrix}.
\end{align}
It should be noted that the last two terms in Eq.~(\ref{eq:GMpot}) vanish under a $\mathbb{Z}_2$ symmetry. However, we retain them in this work.

\par Under ${\rm SU(2)}_L \times {\rm SU(2)}_R$, the fields $\Phi$ and $\Delta$ transform as
\begin{equation}
  \Phi \to U_L \Phi U_R^\dagger, \quad \Delta \to U_L\Delta U_R^\dagger.
\end{equation}
Here, $U_{L,R}$ denotes $\exp(\ii \theta^a_{L,R} \sigma^a)$ for the doublet and $\exp(\ii\theta^a_{L,R} T^a)$ for the triplet, respectively.
The SM electroweak gauge group ${\rm SU(2)}_L \times {\rm U(1)}_Y$ is embedded in the global ${\rm SU(2)}_L \times {\rm SU(2)}_R$ symmetry, with ${\rm SU(2)}_L$ identified by the generators $T_L^i$ and the hypercharge ${\rm U(1)}_Y$ associated with the generator $T_R^3$.

After the EWSB, the neutral scalar fields can be parameterized as
\begin{equation}
  \begin{split}
    {\xi}^0    &\to v_{\xi}+h_{\xi}, \\
    {\phi}^0    &\to \frac{1}{\sqrt{2}}\left(v_{\phi}+h_{\phi}+i a_{\phi}\right) \\
    {\chi}^0    &\to v_{\chi}+ \frac{1}{\sqrt{2}} \left(h_{\chi}+i a_{\chi}\right) .
  \end{split}
\end{equation}
Here, the parameters $v_{\xi}$, $v_{\phi}$ and $v_{\chi}$ represent the VEVs of the corresponding neutral components of these scalar fields. Then the tree level $\rho$ parameter of the GM model can be expressed as
\begin{equation}
  \rho_{\rm tree} \equiv \frac{m^2_W}{m_Z^2 {\cos^2{\theta_W}}}=\frac{v_\phi^2 + 4 v_\chi^2 + 4 v_\xi^2}{v_\phi^2 + 8 v_\chi^2}.
\end{equation}
Therefore, when the triplet VEVs are aligned, $v_\chi = v_\xi \equiv v_\Delta$, the vacuum remains invariant only under the diagonal (custodial) subgroup ${\rm SU(2)}_V \subset {\rm SU(2)}_L \times {\rm SU(2)}_R$ (i.e., when $U_L = U_R$). As a result, the tree-level $\rho$ parameter automatically satisfies $\rho_{\rm tree} = 1$.
In this alignment limit, the VEVs of the scalar fields take the form
\begin{equation}
  \langle\Phi\rangle =  \frac{v_\phi}{\sqrt{2}} {\bf I}_{2\times 2}, \quad \text{and} \quad \langle\Delta\rangle= v_{\Delta} {\bf I}_{3\times 3}.
\end{equation}
The condition for EWSB is given by
\begin{equation}
  v^2=v_\phi^2+8 v_\Delta^2 \simeq \left(246~{\rm GeV}\right)^2.
\end{equation}
The physical scalars are then decomposed into irreducible representations according to the custodial ${\rm SU(2)}_V$ symmetry. The doublet $\Phi$ (interpreted as a ${\bf 2} \otimes {\bf 2}$ representation) decomposes into ${\bf 1} \oplus {\bf 3}$, while the triplet $\Delta$ (as a ${\bf 3} \otimes {\bf 3}$ representation) decomposes into ${\bf 1} \oplus {\bf 3} \oplus {\bf 5}$. One triplet, $G_3$, provides the Nambu-Goldstone bosons eaten by $W$ and $Z$ bosons
\begin{equation}
  G_3^\pm = \cos{\theta_H} \phi^\pm
  + \frac{\sin{\theta_H}}{\sqrt{2}} \left( \chi^\pm + \xi^\pm \right), \quad
  G_3^0= \cos{\theta_H}{a_\phi} + \sin{\theta_H} a_{\chi},
\end{equation}
where the mixing parameter $\theta_H$ is defined by
\begin{eqnarray}
  \tan{\theta_H}\equiv2\sqrt{2}v_\Delta / v_\phi.
\end{eqnarray}
The singlet $H_1^0$ (originating from $\Delta$ field), the triplet ($H_3^{\pm}$ and $H_3^0$), and the quintuplet ($H_5^{\pm\pm}$, $H_5^{\pm}$ and $H_5^0$) are given by
\begin{equation}
  \begin{split}
    H^0_1   &=  \sqrt{\frac{2}{3}} h_{\chi} +\sqrt{\frac{1}{3}} h_{\xi}, \\
    H_3^\pm &=  -\sin{\theta_H}\phi^\pm + \frac{\cos{\theta_H}}{\sqrt{2}} \left(\chi^\pm+\xi^\pm\right) , \quad
    H_3^0   =  \cos{\theta_H} a_{\chi}-\sin{\theta_H} a_{\phi}, \\
    H_5^{\pm\pm}
    &=\chi^{\pm\pm}, \quad
    H_5^{\pm}=\frac{1}{\sqrt{2}} \left( \chi^\pm-\xi^\pm \right), \quad
    H_5^0   =\sqrt{\frac{2}{3}} h_{\xi} -\sqrt{\frac{1}{3}}h_{\chi}.
  \end{split}
\end{equation}
\par In the basis of $(h_\phi, H_1^0, H_5^0)$, the mass matrix of $CP$-even Higgs states is given by
\begin{equation}
  M^2_{CP\text{-even}} =
  \begin{pmatrix}
    M_{11}^2 & M_{12}^2 & 0 \\ M_{12}^2 & M_{22}^2 & 0 \\ 0 & 0 & m_{H_5}^2
  \end{pmatrix},
\end{equation}
where the elements of the $2\times 2$ part are
\begin{equation}
  \begin{split}
    {M}_{11}^2 &=
    8\lambda_1 v_\phi^2,\\
    {M}_{12}^2 ={M}_{21}^2 &=
    \frac{\sqrt{3}}{2} v_\phi \left(4\left(2\lambda_2+\lambda_5\right)v_\Delta+\mu_1\right), \\
    {M}_{22}^2 &=
    8\left(\lambda_3 +4\lambda_4\right)v_\Delta^2
    - \frac{v_\phi^2}{4 v_\Delta} \mu_1
    + 6 v_\Delta \mu_2 .
    \label{eq:singlets mass matrix}
  \end{split}
\end{equation}
The mass eigenstates $h$ and $H$ are defined by
\begin{equation}
  h=\cos{\alpha} h_{\phi} - \sin{\alpha}{H^0_1},\quad
  H=\sin{\alpha} h_{\phi} + \cos{\alpha}{H^0_1},
\end{equation}
where the mixing angle $\alpha$ is determined by
\begin{equation}
  \tan{2\alpha}=\frac{2 M_{12}^2}{M_{22}^2-M_{11}^2}, \quad   \alpha \in \left( -\pi/2, \pi/2 \right)
  \label{eq:singlet mixing angle}
\end{equation}
The masses of all Higgs mass eigenstates can then be expressed as
\begin{equation}
  \begin{split}
    m_{h}^2   &= M_{11}^2 \cos^2{\alpha} + M_{22}^2 \sin^2{\alpha} + M_{12}^2 \sin{2\alpha}, \\
    m_{H}^2  &= M_{11}^2 \sin^2{\alpha} + M_{22}^2 \cos^2{\alpha} - M_{12}^2 \sin{2\alpha}, \\
    m_{H_3}^2 \equiv m_{H_3^{\pm}}^2 = m_{H_3^{0}}^2
    &= -\frac{v^2}{4 v_\Delta} \mu_1 - \frac{1}{2}\lambda_5 v^2,  \\
    m_{H_5}^2 \equiv m_{H_5^{\pm\pm}}^2 = m_{H_5^{\pm}}^2 =m_{H_5^{0}}^2
    &= -\frac{v_\phi^2}{4 v_\Delta} \mu_1 - 12 v_\Delta \mu_2 + 8\lambda_3 v_\Delta^2 - \frac{3}{2}\lambda_5 v_\phi^2.
  \end{split}
\end{equation}

By applying the tadpole conditions,
\begin{equation}
  \frac{\partial V}{\partial h_{\phi} } \bigg|_0 = 0, \quad
  \frac{\partial V}{\partial h_{\xi} } \bigg|_0 = 0, \quad
  \frac{\partial V}{\partial h_{\chi} } \bigg|_0 = 0,
\end{equation}
the mass parameters $m_{\phi}^2$ and $m_{\Delta}^2$ in Eq.~(\ref{eq:GMpot}) can be eliminated as follows:
\begin{equation}
  \begin{split}
    m_\phi^2 &= -4\lambda_1 v_\phi^2 - \left(\frac{3}{2}\mu_1 + 6\lambda_4 + 3\lambda_5\right) v_\Delta^2, \\
    m_\Delta^2 &= - (12\lambda_2 + 4\lambda_3) v_\Delta^2 - \left(\frac{1}{4 v_\Delta} \mu_1 + 2\lambda_4 + \lambda_5\right) v_\phi^2 - 6 v_\Delta \mu_2.
  \end{split}
\end{equation}

\par In summary, the free parameters of the GM model consist of five dimensionless couplings $\lambda_{1,2,3,4,5}$, two dimensional parameters $\mu_{1,2}$, and one mixing parameter $\theta_H$. In practice, we adopt the following set of free parameters in this work:
\begin{equation}\label{eq:params}
  \lambda_{2, 3, 4, 5}, \quad m_h^{\rm In}, \quad m_{H}^{\rm In}, \quad \alpha, \quad \sin{\theta_H},
\end{equation}
where $\mu_{1,2}$ are substituted by $m_{h,H}^{\rm In}$, and $\lambda_1$ is traded for $\alpha$. In the following sections, $m_{h,H}$ without a superscript refers to the physical pole mass as calculated by the spectrum calculator.

%%%%%%%%%%%%%%%%%%%%%%%%%%%%%%%%%%%%%%%%%%%%%%%%%%%%%%%%%%%%
\section{95 GeV excesses in the GM Model}
\label{sec-3}
In the setup introduced in Sec.~\ref{sec-2}, the GM model contains two light singlet scalar fields.
In this work, the heavier state, denoted as $h$, is identified with the $125~\mathrm{GeV}$ SM-like Higgs boson. The other state, labeled as $H$, serves as a candidate to explain the observed excesses around $95~\mathrm{GeV}$.
The couplings of these Higgs states to SM fermions ($f$) and weak gauge bosons ($V = W, Z$) are modified as follows:
\begin{equation}
  \begin{split}
    g_{hff} &= \kappa_{F} \times g_{hff}^{\rm SM}, \quad g_{hVV} = \kappa_{V} \times  g_{hVV}^{\rm SM}, \\
    g_{Hff} &= \kappa_{F}^{H} \times g_{hff}^{\rm SM}, \quad g_{HVV} = \kappa_{V}^{H} \times  g_{hVV}^{\rm SM},
  \end{split}
\end{equation}
with the effective Higgs couplings $\kappa$ in the lowest order
\begin{equation}\label{eq:kappas}
  \begin{split}
    \kappa_{F} = \frac{\cos{\alpha}}{\cos{\theta_H}}, \quad
    \kappa_{V} = \cos{\theta_H} \cos{\alpha} - \sqrt{\frac{8}{3}}\sin{\theta_H} \sin{\alpha}, \\
    \kappa_{F}^{H} = \frac{\sin{\alpha}}{\cos{\theta_H}}, \quad
    \kappa_{V}^{H} = \cos{\theta_H} \sin{\alpha} + \sqrt{\frac{8}{3}}\sin{\theta_H} \cos{\alpha}. \\
  \end{split}
\end{equation}
\par Using this framework, the GM predictions for the di-photon signal strength normalized to the SM prediction $\mu_{\gamma\gamma}$ are given by
\begin{equation}\label{eq:muaa}
  \begin{split}
    \mu_{\gamma\gamma} &\equiv \frac{ \sigma_{\rm GM}\left(pp\to H_{95} \to \gamma\gamma \right)}{ \sigma_{\rm SM}\left(pp \to  h_{95} \to \gamma\gamma \right)} \\
    &\simeq \frac{ \sigma_{\rm GM} \left( gg \to H_{95} \right)} { \sigma_{\rm SM} \left( gg\to h_{95} \right)  } \times \frac{\mathcal{B}_{\rm GM} \left(H_{95} \to \gamma\gamma \right)}{\mathcal{B}_{\rm SM} \left(h_{95} \to \gamma\gamma \right)} \\
    &= \left| \kappa_{F}^{H} \right|^2 \times \frac{\mathcal{B}_{\rm GM} \left(H_{95} \to \gamma\gamma \right)}{1.39 \times 10^{-3}},
  \end{split}
\end{equation}
where the mass of the Higgs boson is fixed around 95 GeV (denoted by $H_{95}$ and $h_{95}$), and the subscript $\rm GM$ ($\rm SM$) denotes the GM (SM) prediction for the inclusive production rate of $H_{95}$ ($h_{95}$). The specific value $1.39\times 10^{-3}$ is the SM Higgs branching ratio at 95 GeV calculated by LHC Higgs Cross Section Working Group~\cite{LHCHiggsCrossSectionWorkingGroup:2013rie}. The second line approximation arises because the production of the SM Higgs boson is dominated by the gluon-gluon fusion process, which accounts for approximately 89\% of the total Higgs production at a mass of 95 GeV~\cite{LHCHiggsCrossSectionWorkingGroup:2013rie}. The third line arises because the gluon-gluon fusion process ($gg \to h$) is mediated by a colored-fermion loop. As a result, the effective coupling $\kappa^{H}_{gg}$ equals $\kappa^{H}_{F}$, since the fermionic couplings of $H$ scale universally as $\kappa_{F}^{H}$ in lowest order, as shown in Eq.~(\ref{eq:kappas}).
Similarly, the GM predictions for the signal strengths in the $\tau\tau$ and $bb$ channels are given by
\begin{equation}\label{eq:mubbtata}
  \begin{split}
    \mu_{bb}     &\equiv  \frac{ \sigma_{\rm GM}\left(e^+ e^- \to Z H_{95} (H_{95} \to bb ) \right)}{ \sigma_{\rm SM}\left(e^+ e^- \to Z h_{95} (h_{95} \to bb ) \right)}  \\
    &=\left| \kappa_{V}^H \right|^2 \times \frac{\mathcal{B}_{\rm GM} \left( H_{95} \to bb \right) }{0.801}, \\
    \mu_{\tau\tau}  &\equiv \frac{ \sigma_{\rm GM}\left(pp\to H_{95} \to \tau\tau \right)}{ \sigma_{\rm SM}\left(pp \to  h_{95} \to \tau\tau \right)}  \\
    &\simeq \left| \kappa_{F}^H \right|^2 \times \frac{\mathcal{B}_{\rm GM} \left( H_{95} \to \tau\tau \right) }{8.32 \times 10^{-2}}.
  \end{split}
\end{equation}
From the above definitions, all signal strengths are expressed in terms of the ratio $\sigma_{\rm GM}/\sigma_{\rm SM}$, so QCD corrections to the numerator and denominator largely cancel. This helps reduce the theoretical uncertainty in the predicted signal strengths.
The LEP $b\bar b$ equality is exact at leading order (Higgsstrahlung $\propto|\kappa_V^H|^2$). The inclusive LHC $\tau\tau$ relation is approximate: it assumes gluon-fusion dominance and neglects smaller $\kappa_V^H$-dependent production modes.
\par Now, we scrutinize the branching ratios. The partial decay width of the lighter scalar $H$ into SM final states can be expressed as
\begin{equation}
  \Gamma (H\to XX) = |\kappa^{H}_{X}|^{2} \times \Gamma_{\rm SM} (h\to XX)\big|_{m_h = m_H},
\end{equation}
where $\Gamma_{\rm SM} (h\to XX)\big|_{m_h = m_H}$ is the SM Higgs partial decay width evaluated at mass $m_{H}$. Therefore, the total decay width of custodial singlet state $H$ can be written as
\begin{equation}\label{eq:HpartialWidth}
  \Gamma^{\rm tot}_{H} = \Gamma^{\rm SM}_{h}  \sum_{X} \left|\kappa^{H}_{X}\right|^{2}  \mathcal{B}_{\rm SM} \left( h \to XX \right),
\end{equation}
where the SM Higgs total decay width $\Gamma^{\rm SM}_{h}$ and branching ratios $\mathcal{B}_{\rm SM} \left( h \to XX \right)$ are evaluated at $m_{H}$.
The couplings between the neutral scalars $H$ and $h$ and photons are induced by charged-particle loops. In the SM, Higgs-boson decay into two photons is mediated by $W$-boson and heavy charged-fermion (top-quark-dominated) loops. In the GM model, loops of heavy charged Higgs states $S_i$ (including $H_{5}^{\pm\pm}$, $H_{5}^{\pm}$, and $H_{3}^{\pm}$) can significantly enhance the partial decay width $\Gamma(H\to \gamma\gamma)$. The corresponding effective coupling $\kappa^{H}_{\gamma\gamma}$ can be expressed as
\begin{equation}
  \begin{small}
    \kappa^{H}_{\gamma} =
    \frac
    {  \sum_{i}\dfrac{g_{HS_iS_i}}{2 m_{S_i}^2} vQ_{S_i}^2 A_0 \left(\frac{m_{H}^2}{4 m_{S_i}^2}\right)
      + \kappa_{V}^{H} A_1 \left(\frac{m_{H}^2}{4 m_{W}^2}\right)
    + \kappa_{F}^{H} ~\dfrac{4}{3} A_{1/2} \left(\frac{m_{H}^2}{4 m_{t}^2} \right) }
    {A_1 \left(\frac{m_{H}^2}{4 m_{W}^2}\right)
    + \dfrac{4}{3} A_{1/2} \left(\frac{m_{H}^2}{4 m_{t}^2} \right)}
    ,
  \end{small}
\end{equation}
where $g_{HS_iS_i}$ are the scalar trilinear couplings of $H$ to the charged states $S_i$, and the loop functions $A_{0, 1/2, 1}$ can be found in the literature~\cite{Djouadi:2005gi, Djouadi:2005gj, Gunion:1989we}. For an $H$ boson with mass $95~{\rm GeV}$, $A_1 \simeq -7.6$ and $A_{1/2} \approx 4/3$. The factor $A_0$ reaches its maximum value of $1.5 + \ii$ at $m_H = 2 m_{S_i}$, while in the limit of very heavy loop charged scalars, $A_0(0) \simeq 1/3$.
\par Then the signal strength in Eq.~(\ref{eq:muaa}) and Eq.~(\ref{eq:mubbtata}) can be rewritten as:
\begin{equation}
  \begin{split}
    \mu_{\gamma\gamma}
    &= \left| \kappa^{H}_{F} \right|^2 \left|  \kappa^{H}_{\gamma} \right|^2 \cdot
    \frac{\Gamma_{h}^{\rm SM}}{\Gamma^{\rm tot}_{H}} \\
    \mu_{bb}
    &= \left| \kappa^{H}_{V} \right|^2  \left| \kappa^{H}_{F} \right|^2 \cdot \frac{\Gamma_{h}^{\rm SM}}{\Gamma^{\rm tot}_{H}}, \\
    \mu_{\tau\tau}
    &= \left| \kappa^{H}_{F} \right|^4  \cdot \frac{\Gamma_{h}^{\rm SM}}{\Gamma^{\rm tot}_{H}}.
  \end{split}
\end{equation}

It is clear that all three signal strengths share a common factor, $\left| \kappa^{H}_{F} \right|^2 \cdot \frac{\Gamma_{h}^{\rm SM}}{\Gamma^{\rm tot}_{H}}$. This introduces tension in attributing all observed excesses to a single source. Charged scalar states in the GM model can provide positive contributions to the $\gamma\gamma$ excess, especially the doubly charged $H_5^{\pm\pm}$ with $Q_S^2 = 4$. Ref.~\cite{Chen:2023bqr} investigates different possibilities for $H$ to explain $\gamma\gamma$, $\gamma\gamma + bb$, and $\gamma\gamma + bb + \tau\tau$.
However, recent dedicated reanalyses argue that this feature is unlikely to originate from a genuine new scalar resonance~\cite{Janot:2024ryq}. Given the inferior di-jet mass resolution at LEP and the updated conclusions, we do not attempt to accommodate the $b\bar b$ hint in this work. Instead, we focus on the two LHC-driven channels ($\gamma\gamma$ and $\tau\tau$) where the experimental systematics and resolutions are better controlled in the relevant mass window.
Finally, we note that all loop-induced rates, branching ratios and total widths in our scan are evaluated with the spectrum tools described in Sec.~\ref{sec-4}.

% ===== SECTION 4 =====

\section{Methodology and Constraints}
\label{sec-4}
In order to realize a light custodial singlet scalar $H$ near $95.4~{\rm GeV}$ that is capable of accommodating the LHC hints in the $\gamma\gamma$ and $\tau\tau$ channels, we perform a comprehensive parameter scan. This section describes the methodology of the scan together with the various constraints imposed. The detailed numerical outcomes corresponding to these constraints will be presented in Sec.~\ref{sec-5}.

In practice, we utilize the package \texttt{GMCALC}~\cite{Hartling:2014xma}, a dedicated GM model calculator. Alternatively, we use the \texttt{SPheno} package~\cite{Porod:2003um, Porod:2011nf} in combination with \texttt{SARAH}~\cite{Staub:2013tta, Staub:2012pb, Staub:2010jh, Staub:2009bi, Staub:2008uz} to compute the GM model spectrum.
\par We perform a comprehensive parameter scan using the \textsc{Jarvis-HEP} framework~\cite{Guo:2026kfy}. The input parameters in Eq.~(\ref{eq:params}) are scanned over the following ranges:
\begin{equation}
  \begin{split}
    -\sqrt{4\pi} &< {\lambda_{2,3,4,5}} <\sqrt{4\pi},\\
    120~\mathrm{GeV} < m_{h}^{\rm In} &< 130~\mathrm{GeV}, \quad
    90~\mathrm{GeV} < m_H^{\rm In} < 102~\mathrm{GeV},
    \\
    0<\sin{\theta_H} &<0.4, \quad \pi/4< |\alpha|<\pi/2.
  \end{split}
\end{equation}

To assess the global viability of the GM model in interpreting the observed excesses, we construct a likelihood-based framework. The total likelihood $\mathcal{L}_{\rm tot}$ is defined as
\begin{equation}
  \mathcal{L}_{\rm tot} \,=\, \exp\!\left(-\frac{1}{2}\,\chi^2_{\rm tot}\right)
  \times \Theta_{\rm PD}\,\Theta_{\rm unit}\,\Theta_{\tt HB}\,\Theta_{\tt HS},
  \qquad
  \chi^2_{\rm tot}
  = \chi^2_{\rm mass}
  + \chi^2_{95}
  + \chi^2_{\rm EW}
  + \chi^2_{\kappa}
  + \chi^2_{B}.
  \label{eq:likelihood}
\end{equation}
The total $\chi^2_{\rm tot}$ contains only genuine likelihood contributions from the experimental observables, while the factors $\Theta_{\rm PD}$, $\Theta_{\rm unit}$, $\Theta_{\tt HB}$, and $\Theta_{\tt HS}$ act as acceptance criteria that restrict the likelihood support to theoretically and phenomenologically viable regions of parameter space.

The individual $\chi^2$ and $\Theta$ terms entering Eq.~\eqref{eq:likelihood} are defined as follows.
\begin{itemize}
  \item $\chi^2_{\rm mass} = \left( \dfrac{m_h - 125.18~\rm{GeV}}{1~\rm{GeV}} \right)^2 + \left( \dfrac{m_H - 95.4~\rm{GeV}}{1~\rm{GeV}} \right)^2$.
    \par Here, $m_h$ and $m_H$ are the theoretical predictions of the two $CP$-even Higgs states, and the total uncertainties (theoretical and experimental) are both set to $1~{\rm GeV}$.

  \item $\Theta_{\tt HB}$ implements the direct-search constraints from \texttt{HiggsBounds-5.10.0}~\cite{Bahl:2022igd, Bechtle:2008jh, Bechtle:2011sb, Bechtle:2013wla, Bechtle:2020pkv}. We define $\Theta_{\tt HB}=1$ for parameter points that satisfy the HiggsBounds exclusion test and $\Theta_{\tt HB}=0$ otherwise. In this way, direct-search constraints act as a veto on the parameter space rather than as a compensating contribution to the global $\chi^2$.

  \item $\Theta_{\tt HS}$ encodes the consistency requirement from \texttt{HiggsSignals-2.6.2}~\cite{Bechtle:2013xfa, Bechtle:2014ewa, Bechtle:2020uwn, Bahl:2022igd}. To avoid double counting Higgs signal-strength information with the explicit $\chi^2_{\kappa}$ term defined below, we do not include the HiggsSignals output as an additive likelihood contribution in $\chi^2_{\rm tot}$. Instead, we require that the sample satisfy $\chi^2({\rm GM}) < \chi^2({\rm SM})$, where $\chi^2({\rm SM})\approx 152.5$ denotes the SM reference value returned by \texttt{HiggsSignals} for the 125~GeV Higgs rate measurements in our setup. We therefore set $\Theta_{\tt HS}=1$ when this condition is fulfilled and $\Theta_{\tt HS}=0$ otherwise.

  \item $\chi^2_{95} = \left( \dfrac{\mu_{\gamma\gamma} - 0.24}{0.09} \right)^2 + \left( \dfrac{\mu_{\tau\tau} - 1.2}{0.5} \right)^2$: denotes the two $95~{\rm GeV}$ excesses, where $\mu_{\gamma\gamma}$ and $\mu_{\tau\tau}$ are the theoretical predictions including loop corrections.
  \item The contribution $\chi^2_{\rm EW}$ encodes constraints from electroweak precision measurements in terms of the oblique parameters $S$, $T$, and $U$, which capture the one-loop contributions of the additional scalar states to gauge-boson self-energies and neutral-current processes~\cite{Peskin:1990zt, Peskin:1991sw}. In this work, the $\chi^2_{\rm EW}$ is implemented using a multivariate Gaussian form,
    \begin{equation}
      \chi^2_{\rm EW}    = \Delta\vec{O}_{\rm EW}^{\,T} \,\rho_{\rm EW}^{-1}\, \Delta\vec{O}_{\rm EW},
    \end{equation}
    where
    \begin{equation}
      \Delta\vec{O}_{\rm EW}^T = \left(\frac{S + 0.04}{0.10},\; \frac{T - 0.01}{0.12}, \; \frac{U-0.05}{0.09}\right)
    \end{equation}
    and $\rho_{\rm EW}$ represents the correlation matrix, defined as
    \begin{equation}
      \rho_{\rm EW} =
      \begin{pmatrix}
        1       &     0.93    & -0.70 \\[4pt]
        0.93        &   1       & -0.87 \\[4pt]
        -0.70   &   -0.87  & 1
      \end{pmatrix}.
    \end{equation}
    In addition, the $\rho$ parameter, which characterizes possible new sources of ${\rm SU(2)}$ breaking beyond the SM Higgs doublet, is constrained to be
    \begin{equation}
      \rho = 1.00031 \pm 0.00019.
    \end{equation}
    In the custodial-symmetric GM setup, $\rho_{\rm tree}=1$ holds by construction, and any residual deviation arises only at loop level.
  \item The contribution $\chi^2_{\kappa}$ encodes the constraints on the Higgs coupling modifiers through a two-dimensional Gaussian likelihood in the $(\kappa_V,\kappa_F)$ plane, and is defined as
    \begin{equation}
      \chi^2_{\kappa} = \Delta\vec{\kappa}^{\,T}\, V^{-1}\, \Delta\vec{\kappa},
    \end{equation}
    where
    \begin{equation} \Delta\vec{\kappa} =
      \begin{pmatrix}
        \kappa_V - \hat\kappa_V \\
        \kappa_F - \hat\kappa_F
      \end{pmatrix},
    \end{equation}
    and $V$ denotes the corresponding covariance matrix,
    \begin{equation}
      V =
      \begin{pmatrix}
        \sigma_V^2 & 0.39\,\sigma_V\sigma_F \\
        0.39\,\sigma_V\sigma_F & \sigma_F^2
      \end{pmatrix}.
    \end{equation}
    In our analysis, we adopt the ATLAS combined fit results shown in Fig.~4 of Ref.~\cite{ATLAS:2022vkf}, with $\hat\kappa_V = 1.035 \pm 0.031$ and $\hat\kappa_F = 0.95 \pm 0.05$, and a correlation coefficient of $0.39$.
  \item $\chi_B^2 = \left( \dfrac{\mathcal{B}(B_s\to \mu^+ \mu^-) - 3.1\times{10^{-9}}}{1.3\times{10^{-9}}} \right)^2 + \left( \dfrac{\mathcal{B}(B\to X_s \gamma) - 3.43\times{10^{-4}}}{0.44\times{10^{-4}}} \right)^2$ accounts for constraints from rare $B$-meson decays, which are strongly suppressed in the SM and therefore provide sensitive probes of physics beyond the SM, in particular extended Higgs sectors~\cite{CMS:2014xfa}.
\end{itemize}

\par In addition to the experimental limits mentioned above, theoretical considerations further constrain the parameter space through the Heaviside step functions $\Theta_{\rm unit}$ and $\Theta_{\rm PD}$ in Eq.~(\ref{eq:likelihood}). These functions act as strict vetoes, restricting the likelihood support to regions of parameter space that are theoretically consistent, where $\Theta(x)$ equals one for $x \ge 0$ and zero otherwise.

\par The $\Theta_{\rm unit}$ enforces perturbative unitarity of scalar $2\to2$ scattering, requiring that the zeroth partial-wave amplitudes obey $|{\rm Re}\,a_0|<\tfrac{1}{2}$ (equivalently $|a_0|<1$). In the GM model this translates into bounds on linear combinations of quartic couplings~\cite{Hartling:2014zca,Chowdhury:2024mfu,Aoki:2007ah}, leading to the conditions
\begin{equation}\label{eqn:pertUnitarity}
  \begin{split}
    \sqrt{(6 \lambda_1 -7 \lambda_3 -11 \lambda_2)^2 +36 \lambda_4^2} + \left| 6 \lambda_1 + 7 \lambda_3 +11 \lambda_2 \right| &<  4 \pi, \\
    \sqrt{(2 \lambda_1 + \lambda_3 -2 \lambda_2)^2 + \lambda_5^2} + \left| 2 \lambda_1 - \lambda_3 +2 \lambda_2 \right| &< 4 \pi, \\
    \left| 2 \lambda_3 + \lambda_2 \right| < \pi,  \qquad
    \left| \lambda_4 + \lambda_5 \right| &< 2 \pi.
  \end{split}
\end{equation}
Therefore, $\Theta_{\rm unit} = 1$ if all conditions in Eq.~(\ref{eqn:pertUnitarity}) are satisfied simultaneously.

\par The positive-definiteness requirement is imposed to ensure the stability of the electroweak vacuum, such that the scalar potential remains bounded from below for large field values. The quantum corrections can significantly modify the shape of the scalar potential, potentially leading to violations of the tree-level bounded-from-below (BFB) conditions~\cite{Hartling:2014zca}. Such effects may induce deeper vacua at large field values or render the potential unbounded, implying that the electroweak vacuum preserving custodial symmetry may not be the true ground state. This situation is reminiscent of the metastability of the SM electroweak vacuum for a $125~{\rm GeV}$ Higgs boson.
\par To address this issue, we impose the positive definiteness constraints~\cite{Du:2024mry} to guarantee the stability of the custodial vacuum by requiring positivity of certain subdeterminants of the quartic coupling matrices in the one-loop RG-improved potential.
For simplicity, the one-loop RG-improved scalar potential is parameterized by the dimensionless couplings $\lambda$, $\rho_{1,2,\dotsc,5}$, and $\sigma_{1,2,3,4}$ of the most general ${\rm SU(3)}_C \times {\rm SU(2)}_L \times {\rm U(1)}_Y$ invariant GM model.\footnote{Dimensional parameters are eliminated via the minimization conditions for the VEVs, and thus do not affect the large-field analysis relevant to positive definiteness.}
In the custodial-symmetric limit these couplings map onto the standard GM quartics $\lambda_{1,2,\dots,5}$ as
\begin{equation}\label{rel}
  \begin{split}
    \lambda  &=4\lambda_1, \\
    \rho_1  &=4\lambda_2+6\lambda_3,\quad
    \rho_2  =-4\lambda_3,\quad
    \rho_3  =2(\lambda_2+\lambda_3),\quad
    \rho_4  =4\lambda_2,\quad
    \rho_5  =4\lambda_3, \\
    \sigma_1 &= 4\lambda_4-\lambda_5,\quad
    \sigma_2 =2\lambda_5,\quad
    \sigma_3 =2\lambda_4,\quad
    \sigma_4 =\sqrt{2}\lambda_5.
  \end{split}
\end{equation}
With these definitions and when the $\rho_5$ and $\sigma_4$ are small, the positive definiteness constraints then reduce to requiring that all leading principal minors of $\mathcal{V}$ are strictly positive,
\begin{equation}
  \det \mathcal{V}_n > 0 \qquad (n=1,2,\dotsc,5),
\end{equation}
where $\mathcal{V}$ is defined as
\begin{equation}
  \mathcal{V} =
  \begin{pmatrix}
    \frac{1}{4}\lambda   & \frac{1}{8}(2\sigma_1+\sigma_2)    &\frac{1}{4} \sigma_1   &\frac{1}{2}\sigma_3 ~&\frac{1}{4} \sigma_3 \\[4pt]
    \frac{1}{8}(2\sigma_1+\sigma_2)   &\frac{1}{2}(2\rho_1+\rho_2  )&\frac{3}{2}\rho_2    &\rho_4    &\frac{1}{2}\rho_4\\[4pt]
    \frac{1}{4} \sigma_1  &\frac{3}{2} \rho_2   &\rho_1+\rho_2    &\rho_4        &\frac{1}{2}\rho_4 \\[4pt]
    \frac{1}{4}\sigma_3  &\rho_4        &\rho_4        &2\rho_3        &\rho_3\\[4pt]
    \frac{1}{2}\sigma_3  &\frac{1}{2}\rho_4  &\frac{1}{2}\rho_4   &\rho_3        &\frac{1}{2}\rho_3
  \end{pmatrix},
\end{equation}
and $\mathcal{V}_n$ denotes the $n \times n$ upper-left submatrix of $\mathcal{V}$. Therefore, the $\Theta_{\rm PD}$ in Eq.~(\ref{eq:likelihood}) is defined as
\begin{equation}
  \Theta_{\rm PD} = \prod_{n} \Theta(\det \mathcal{V}_n).
\end{equation}
In practice, the full matrix is used numerically,
while the analytic form shown here corresponds
to the simplified limit.
\par It should be noted that the coupling parameters (such as $\lambda$, $\rho_{1,2,\dotsc,5}$, and $\sigma_{1,2,3,4}$, etc.) must be replaced by their renormalization group evolved values in regions of large field values. These evolved parameters can then be used to determine the vacuum structure of the GM model.
We utilize the iterative method described in~\cite{Du:2024mry} to assess whether the parameter space that adequately explains the observed excess signals satisfies the positive definiteness constraints. The gauge coupling values at the EW scale employed in this study are derived from~\cite{Antusch:2013jca}. To ensure that the vacuum maintains custodial symmetry, the boundary conditions for the scalar quartic couplings at the EW scale, which are crucial for the RGEs, are selected in accordance with Eq.~(\ref{rel}).
To determine the scale dependence of the quartic couplings within the one-loop RG-improved tree-level scalar potential, it is essential to evolve the input parameters to an arbitrary scale, especially in regions of large field values, using the appropriate beta functions. In our detailed numerical analysis, to establish the positive definiteness constraints, we evolve the relevant quartic couplings and mass dimension parameters from the EW scale ($M_Z=91.2~{\rm GeV}$) up to the Planck scale ($M_{\rm Planck}\approx 1.12\times{10}^{19}~{\rm GeV}$) or to the lowest Landau pole scale, where any of the quartic couplings diverge.

\par Tree-level BFB and the RG-improved positive-definiteness condition are not pointwise nested: the former tests the custodial-symmetric quartic potential at the electroweak scale, while the latter follows the general gauge-invariant potential under one-loop evolution to the Planck scale or the first Landau pole. As shown in Refs.~\cite{Du:2024mry,Du:2025eop}, positive-definiteness condition both excludes some BFB-allowed regions and admits previously excluded ones, yielding a net expansion of the viable space for the $95~\mathrm{GeV}$ excesses.

To ensure a numerically reliable assessment of the GM model's ability to simultaneously explain the two observed excesses, we conducted an extensive exploration of the parameter space using a combination of random scans, Markov Chain Monte Carlo (MCMC) techniques, and MultiNest sampling. These complementary methods provide both broad global coverage and efficient identification of phenomenologically relevant regions with higher likelihood. The final profile-likelihood sample contains $\mathcal{O}(10^{8})$ retained points after all selections. Convergence was confirmed by verifying the stability of the profile likelihood (PL) with additional sampling.

%%%%%%%%%%%%%%%%%%%%%%%%%%%%%%%%%%%%%%%%%%%%%%%
% Section 5: Numerical results
\section{Numerical results}
\label{sec-5}
In this section, we present detailed results from our sampling. For clarity, we have filtered out all samples that fail the \texttt{HiggsBounds} or \texttt{HiggsSignals} consistency requirements, the positive-definiteness requirement, or the perturbative unitarity constraints; thus, all samples shown in the figures satisfy $\Theta_{\tt HB}=1$, $\Theta_{\tt HS}=1$, $\Theta_{\rm PD}=1$, and $\Theta_{\rm unit}=1$.

\subsection{Fit results and benchmarks}
\begin{figure}
  \centering
  \includegraphics[width=.72\textwidth]{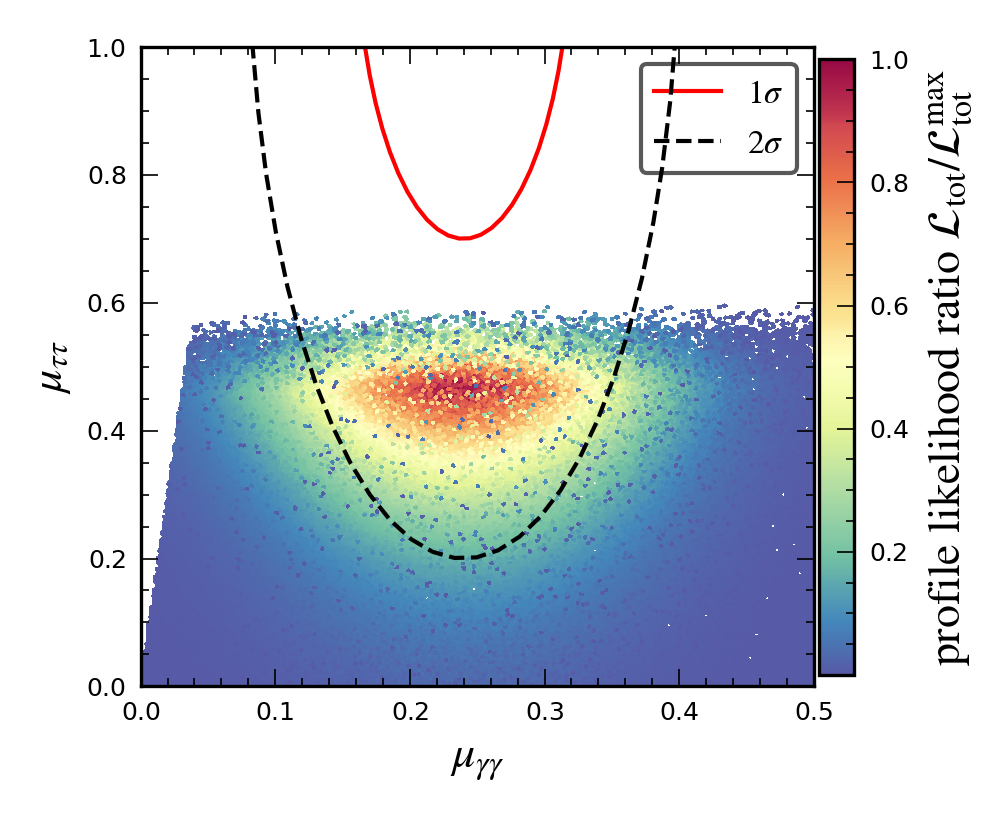}
  \caption{\label{fig:excesses}Profile likelihood map in the $(\mu_{\gamma\gamma}, \mu_{\tau\tau})$ plane. Solid and dashed curves show the experimental $1\sigma$ and $2\sigma$ contours from $\chi^2_{95}$.}
\end{figure}

Figure~\ref{fig:excesses} shows the profile likelihood in the $(\mu_{\gamma\gamma},\,\mu_{\tau\tau})$ plane for the $95~\mathrm{GeV}$ scalar after the constraints of Sec.~\ref{sec-4}. The preferred region lies near $\mu_{\gamma\gamma}\sim0.25$ and $\mu_{\tau\tau}\sim0.47$: the di-photon excess is accommodated near its central value, while the di-tau excess is only partially reproduced and remains within the experimental $2\sigma$ contour.

\par Compared with Ref.~\cite{Chen:2023bqr}, where the tree-level BFB analysis populates mainly $\mu_{\tau\tau}\lesssim0.1$, the RG-improved treatment yields a characteristic di-tau strength larger by a factor of roughly four to five. Different scans and likelihoods mean this is a phenomenological cross-check rather than an isolated measure of the vacuum-stability effect.

\begin{figure}
  \centering
  \makebox[\textwidth][c]{%
    \includegraphics[width=.35\textwidth]{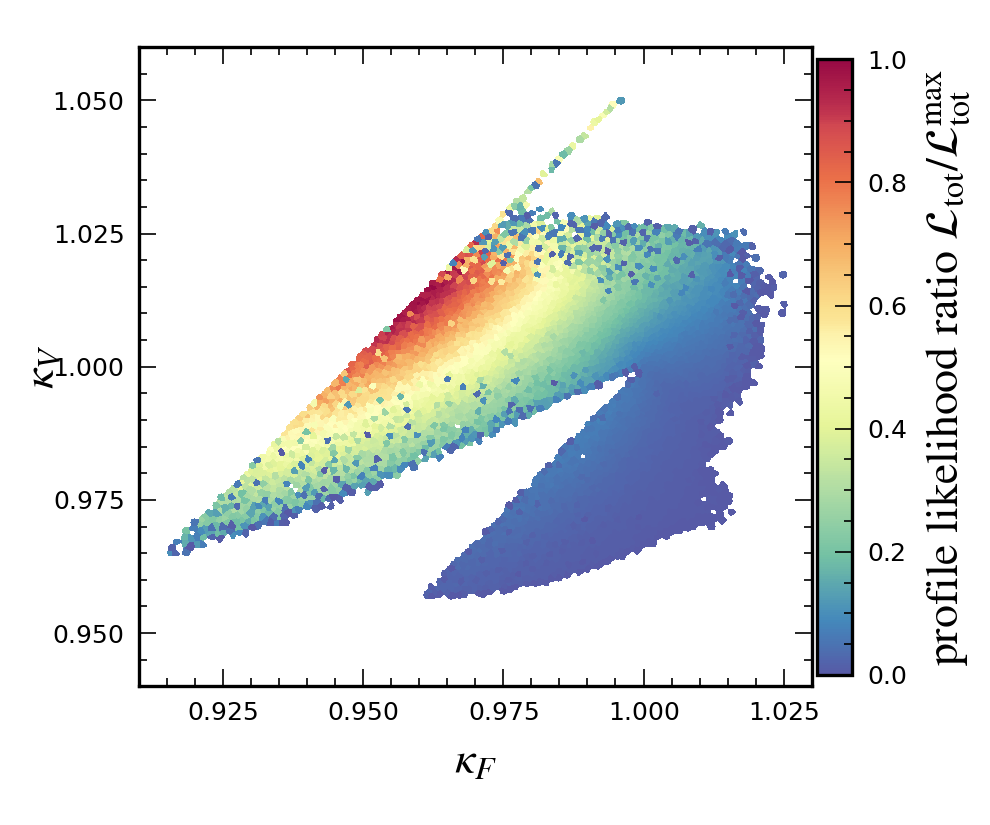}
    \includegraphics[width=.35\textwidth]{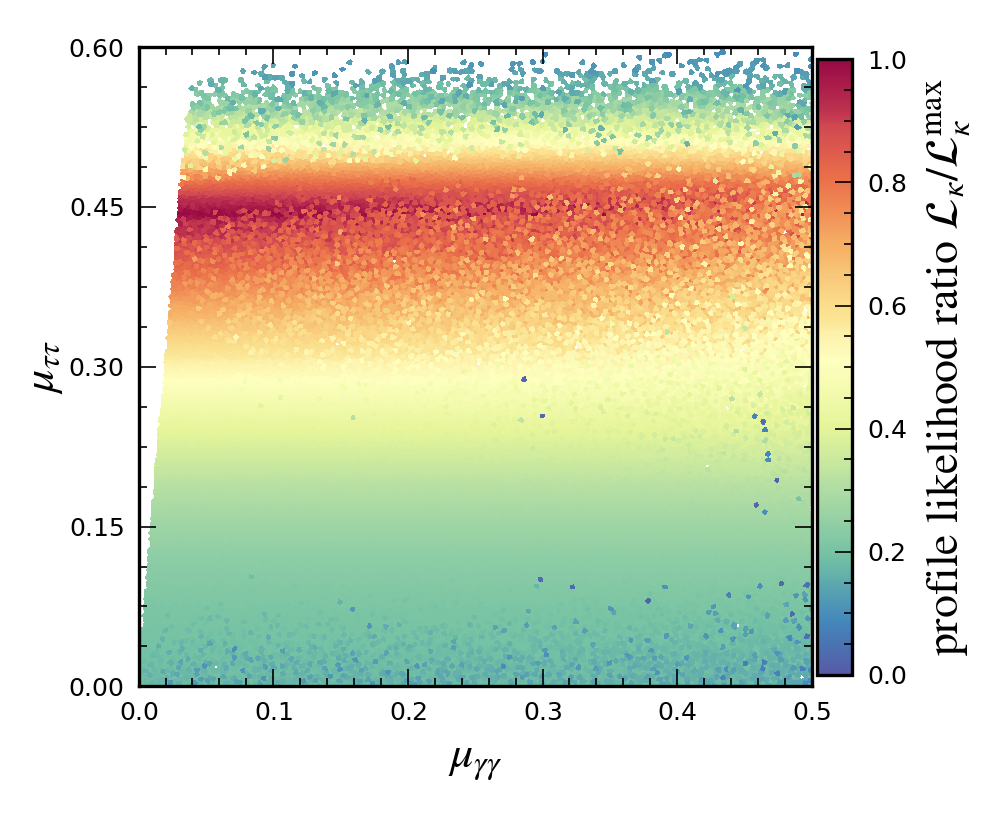}
    \includegraphics[width=.35\textwidth]{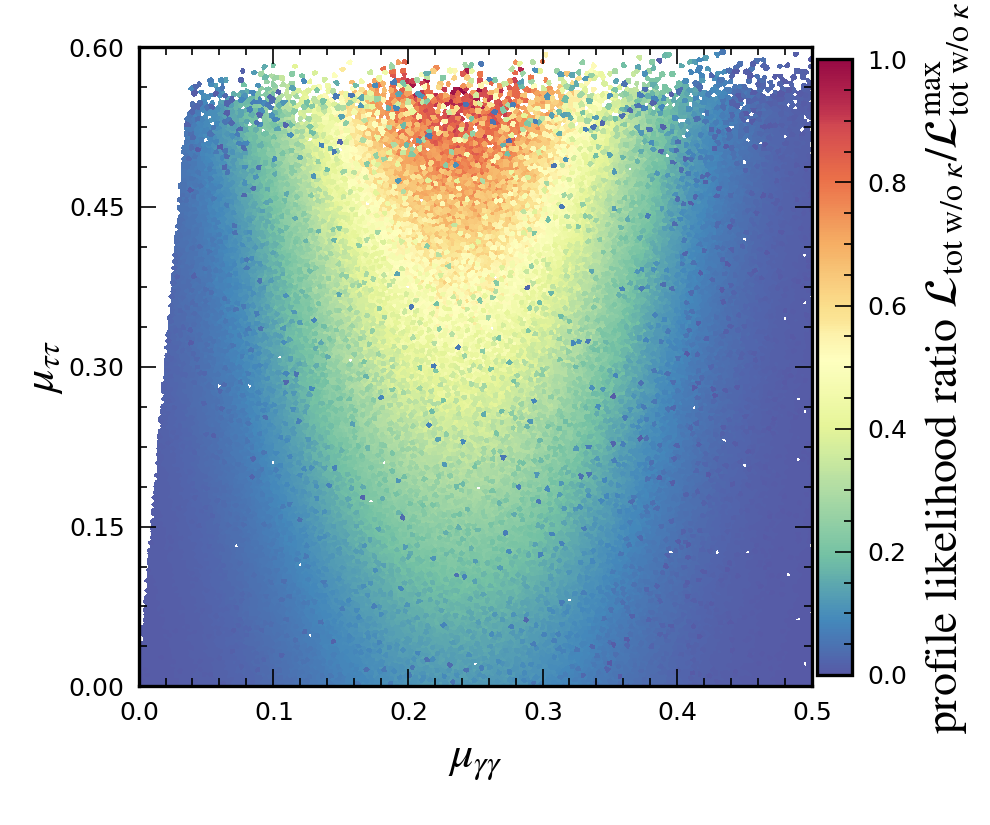}
  }
  \caption{
    \label{fig:kps}
    {\it Left}: The PL map on the $(\kappa_F, \kappa_V)$ plane.
    {\it Middle}: The PL $\mathcal{L}_\kappa$ on the $(\mu_{\gamma\gamma}, \mu_{\tau\tau})$ plane.
    {\it Right}: The PL $\mathcal{L}_{\rm tot}$ without $\mathcal{L}_\kappa$ contribution on the $(\mu_{\gamma\gamma}, \mu_{\tau\tau})$ plane.
  }
\end{figure}

\par The incomplete accommodation of the di-tau excess follows from the $125~\mathrm{GeV}$ Higgs-coupling constraints: $\mu_{\tau\tau}$ cannot be raised independently of $(\kappa_F,\kappa_V)$, which share the mixing angles $\alpha$ and $\theta_H$, as shown in Eqs.~(\ref{eq:mubbtata}) and~(\ref{eq:kappas}). Fig.~\ref{fig:kps} (left) shows the resulting narrow $(\kappa_F,\kappa_V)$ band; $\mathcal{L}_\kappa$ alone favors $\mu_{\tau\tau}\sim0.4$ (middle), while the remaining likelihoods prefer larger values (right). Increasing $|\kappa_F^H|$ also enhances gluon-fusion production and direct-search exposure. These competing effects fix the combined optimum near the preferred region of Fig.~\ref{fig:excesses} and prevent the fit from approaching the CMS central value.

%%%%%%%%%%%%%%%%%%%%%%%%%%%%%%%%%%%%%%%%%%%%%%%%%%%%%%%%%%%%%%%%
%%%%%    Table
\begin{table}[htbp]
  \centering
  \caption{Benchmark points in the GM model.
    The $\chi^2_{\rm tot}$ is defined in Eq.~(\ref{eq:likelihood}).
    A checkmark (\cmark) and a cross (\xmark) indicate whether a given
  benchmark point passes or fails the corresponding constraint. All dimensional quantities are given in units of GeV.}
  \label{tab:Benchmarks}
  \setlength{\tabcolsep}{3.5pt}
  \renewcommand{\arraystretch}{1.15}
  \resizebox{\textwidth}{!}{%
    \begin{tabular}{r S[table-format=3.3] S[table-format=3.3] S[table-format=3.3] c|
        r S[table-format=3.3] S[table-format=3.3] S[table-format=3.3] c|
      r S[table-format=3.3] S[table-format=3.3] S[table-format=3.3]}
      \toprule \hline
      & {\tt BP1} & {\tt BP2} & {\tt BP3}
      & && {\tt BP1} & {\tt BP2} & {\tt BP3}
      & && {\tt BP1} & {\tt BP2} & {\tt BP3} \\
      \hline
      $\lambda_2$        & -0.8211 & -0.6069    &  0.0698
      && $m_H$            &  96.8491 &  96.68 &  95.22
      && $\mu_{\gamma\gamma}$ & 0.2899 & 0.2368 & 0.2445 \\

      $\lambda_3$        &  0.8150 &  0.8336   & -0.8805
      && $m_h$            & 124.0331 & 125.2977 & 125.0191
      && $\mu_{\tau\tau}$     & 0.7423 & 0.5245 & 0.5445 \\

      $\lambda_4$        &  0.5232 & -0.0417 &  0.4704
      && $m_{H_3}$        &  194.5481 & 214.3225 & 222.6281
      && $\chi^2_{\rm tot}$        & 19.795 & 21.220 & 7.605 \\

      $\lambda_5$        &  0.2216 &  0.8814 & -0.0418
      && $m_{H_5}$        & 290.7393 & 340.3623 & 311.1941
      && $\mathcal{B}(H_{5}^{\pm\pm}\!\to\!W^\pm W^\pm)$        & 0.1757 & 0.0032 & 0.1059 \\

      $\alpha$             & -0.8075  &  1.0548  & -1.0399
      && $\kappa_{F}^h$    & 0.8538    & 0.9328  & 0.9330
      && $\mathcal{B}(H_{5}^{\pm\pm}\!\to\! H_3^\pm W^\pm)$      & 0.8243 & 0.9968 & 0.8941 \\

      $\sin\theta_H$       &  0.1309   &  0.0287   &  0.1246
      && $\kappa_{V}^h$    & 0.9296    & 0.9200  & 0.9836
      && Unitarity      & {\cmark}   & {\cmark}   & {\cmark} \\

      $M_H^{\rm in}$     &  96.84 &  96.6780 &  95.2205
      && $S$              & -0.0832 & -0.0930 & -0.0850
      && Positive definiteness & {\cmark} & {\cmark} & {\cmark}\\

      $M_h^{\rm in}$     & 124.0331 & 125.2977 & 125.0191
      && $T$              & -0.0136 & -0.0012 & -0.0181
      && \texttt{HiggsSignals} & {\xmark}   & {\cmark} & {\cmark} \\

      &  &   &
      && $U$                & 0.0007    & 0.0003  & 0.0004
      && \texttt{HiggsBounds}  &  {\cmark} & {\cmark}  & {\cmark}\\
      \hline \bottomrule
  \end{tabular}}
\end{table}

\par To highlight the characteristic features of this surviving region, we present three benchmark points in Table.~\ref{tab:Benchmarks}. \texttt{BP1} features a relatively large triplet admixture, with $\sin\theta_H = 0.131$ and $\alpha = -0.808$. This results in a significant suppression of the SM-like Higgs coupling $\kappa_F$. Consequently, the 95~GeV scalar attains a sizable $\mu_{\tau\tau} = 0.742$ and the highest di-photon rate among the three points, $\mu_{\gamma\gamma} = 0.290$, leading to the smallest $\chi^2_{\rm 95}$. However, it does not satisfy the \texttt{HiggsSignals} constraints. \texttt{BP2} represents the opposite limit with a very small triplet VEV, $\sin\theta_H = 0.029$. The 125~GeV Higgs is therefore more SM-like. The reduced mixing suppresses the $\tau\tau$ signal strength to $\mu_{\tau\tau}=0.525$, while the di-photon rate, $\mu_{\gamma\gamma}=0.237$, is mainly maintained by charged-scalar loop effects. This benchmark satisfies both \texttt{HiggsSignals} and \texttt{HiggsBounds}. Compared to \texttt{BP1}, \texttt{BP3} provides a better solution with $\sin\theta_H = 0.125$ and $\alpha=-1.040$, for which $\kappa_V^h = 0.984$ is remarkably close to experimental data. This point achieves a balanced configuration, $\mu_{\tau\tau}=0.545$ and $\mu_{\gamma\gamma}=0.245$, while remaining consistent with all theoretical and experimental constraints. It exemplifies the thin, correlated region identified in Fig.~\ref{fig:kps}, where the 95~GeV excesses can be accommodated without significantly disturbing the 125~GeV Higgs couplings.

%%%%%%%%%%%%%%%%%%%%%%%%%%%%%%%%%%%%%%%%%%%%%%%%%%%%%
\begin{figure}[th]
  \centering
  \includegraphics[width=.49\linewidth]{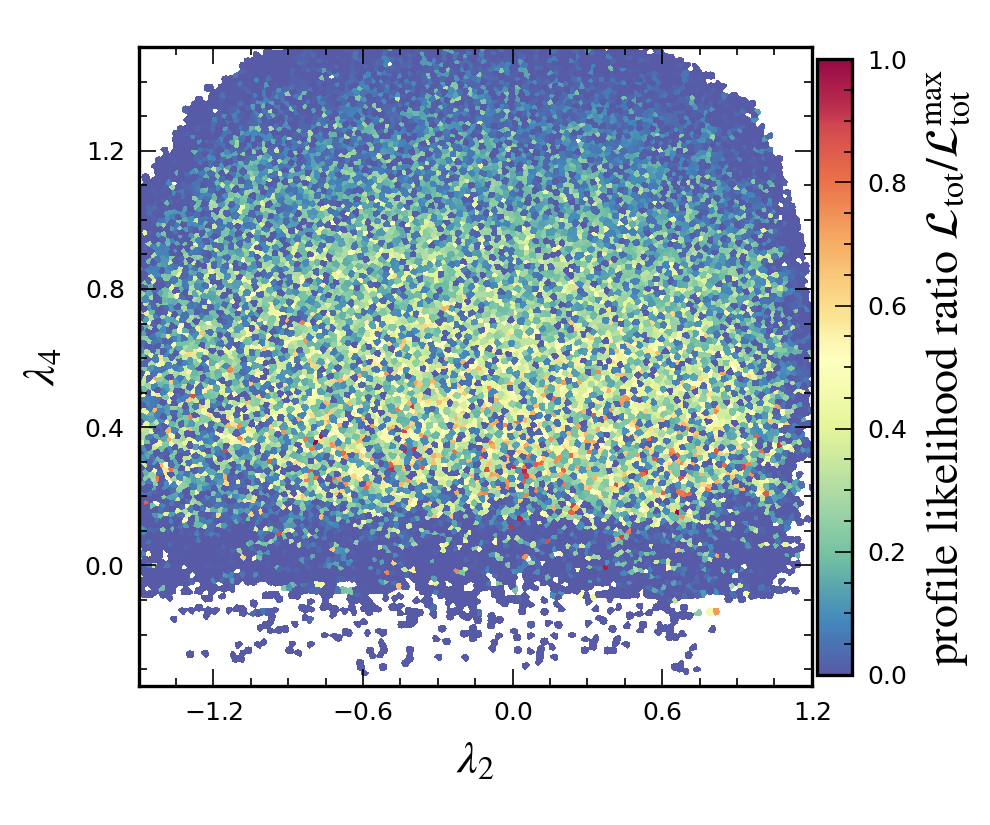}
  \includegraphics[width=.49\linewidth]{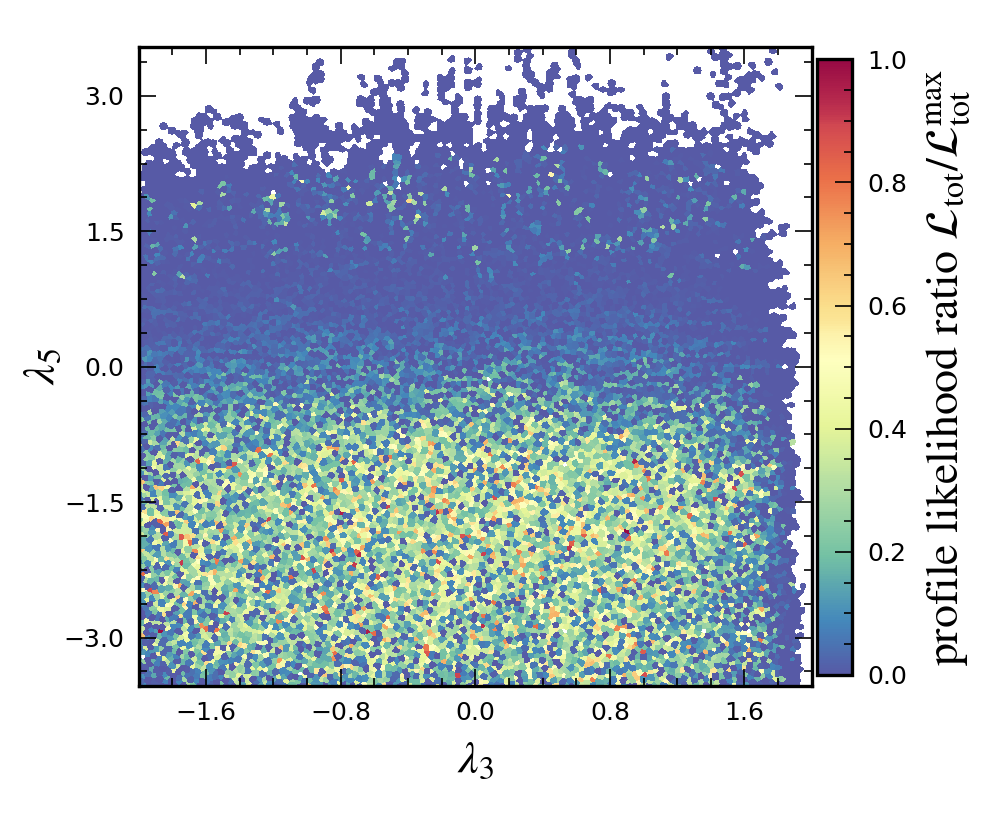} \\\vspace{-.3cm}
  \includegraphics[width=.49\linewidth]{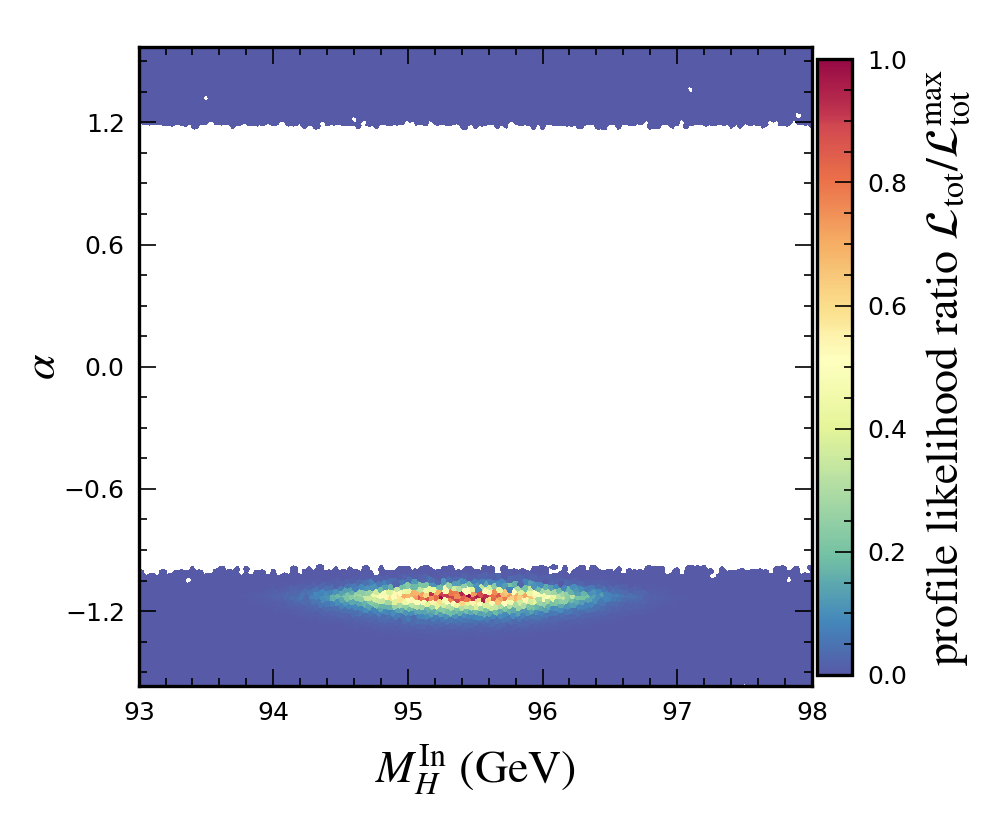}
  \includegraphics[width=.49\linewidth]{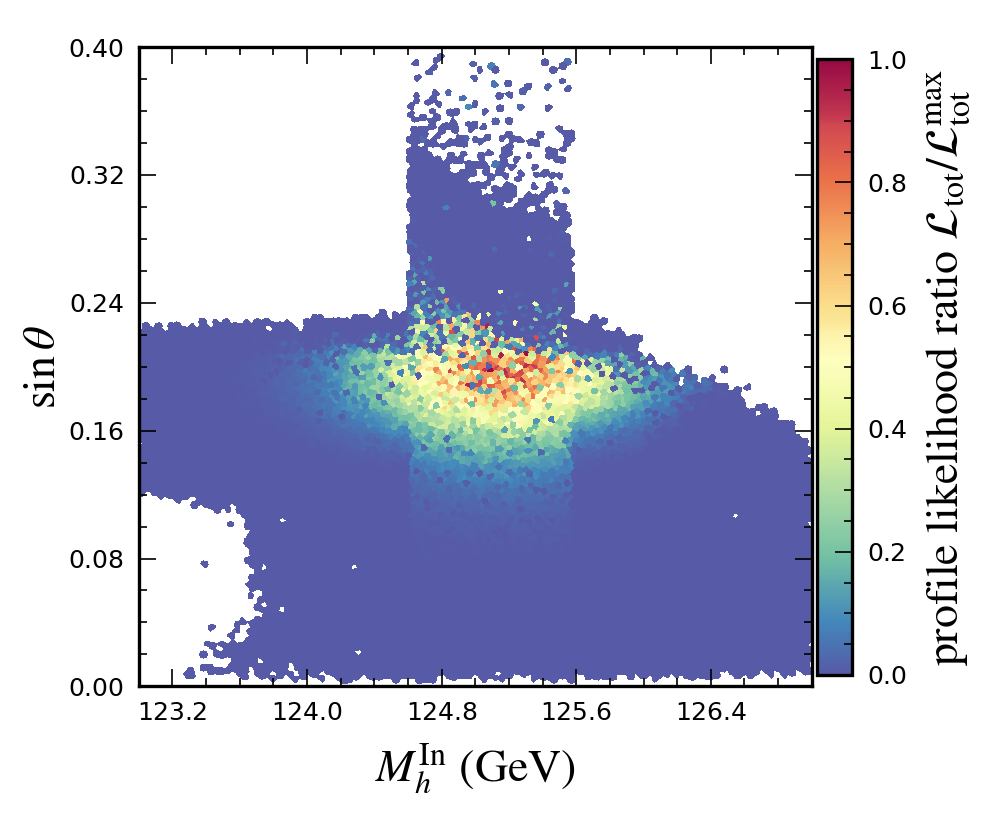} \vspace{-.5cm}
  \caption{\label{fig:params} Similar to Fig.~\ref{fig:excesses}, but projected onto the GM model parameter planes.}
\end{figure}
%%%%%%%%%%%%%%%%%%%%%%%%%%%%%%%%%%%%%%%%%%%%%%%%%%%%%

\subsection{Parameter space and phenomenology}

\par Figure~\ref{fig:params} displays the PL projected onto the eight GM model parameter planes. A clear pattern emerges: while the quartic couplings $\lambda_2$, $\lambda_3$, $\lambda_4$, and $\lambda_5$ span relatively broad ranges, the physically relevant mixing parameters are tightly constrained. In the two $\lambda_i$ panels, no pronounced minima appear, indicating that the quartic interactions are not individually strongly constrained. Instead, these couplings primarily support the scalar mass spectrum and enable viable loop contributions, without requiring fine-tuned correlations among themselves. In contrast, the $(M_H^{\rm in},\alpha)$ projection reveals a highly concentrated high-likelihood region centered at $(95~{\rm GeV}, -\frac{\pi}{3})$, directly corresponding to the mass of the observed excess. The narrow range of the mixing angle $\alpha$ is consistent with the data, and large portions of the $\alpha$ parameter space are excluded, indicating that doublet–triplet mixing is tightly constrained.
For values of $\alpha$ with identical magnitude as shown in the benchmark points \texttt{BP2} and \texttt{BP3}, $\kappa_F$ remains unchanged, as expected. However, the coupling $\kappa_V$ associated with positive $\alpha$ is significantly smaller than that for negative $\alpha$, resulting in a pronounced deviation from experimental constraints. Moreover, $\kappa_V^H$ corresponding to positive $\alpha$ is substantially larger than its counterpart for negative $\alpha$, rendering this scenario unable to survive the additional Higgs-search constraints in the \texttt{HiggsBounds} package.
This feature underlies the $\kappa$ tension observed in Fig.~\ref{fig:kps}: while $\alpha$ at $-\frac{\pi}{4}$ would better explain the excess, the overall fit only allows $\alpha$ near $-\frac{\pi}{3}$. A similar pattern appears in the $(M_h^{\rm In},\sin\theta_H)$ plane. Overall, Fig.~\ref{fig:params} demonstrates that the triplet admixture must remain moderate.

\begin{figure}
  \centering
  \includegraphics[width=0.49\linewidth]{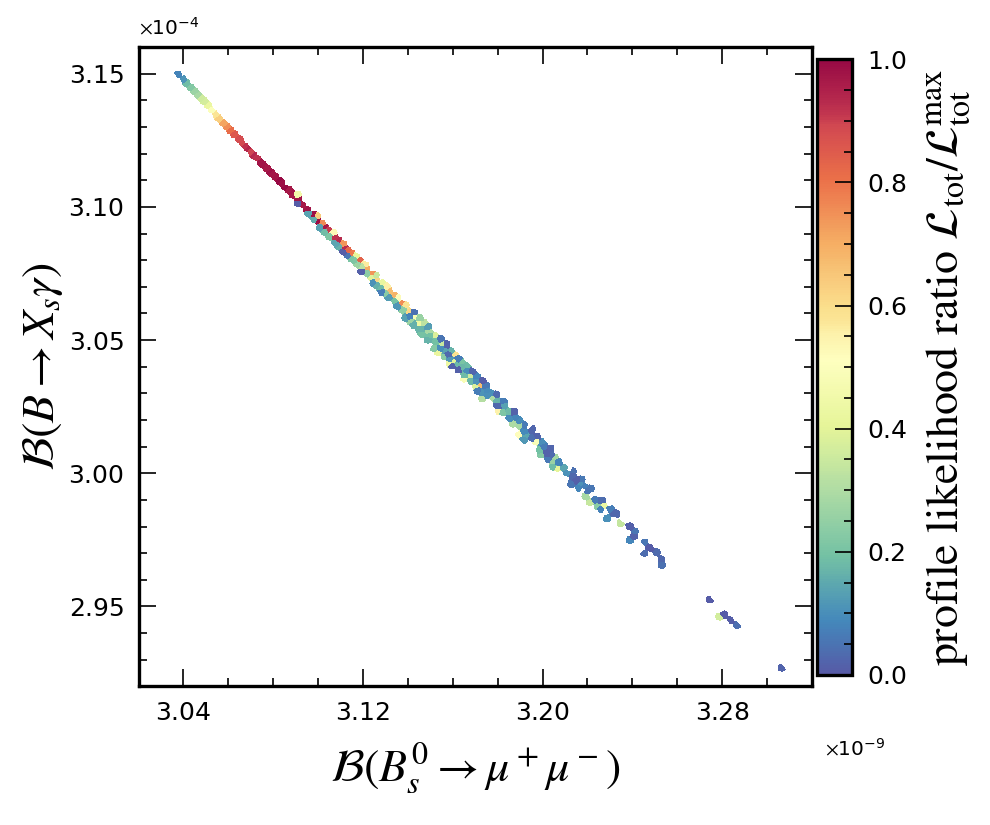}
  \includegraphics[width=0.49\linewidth]{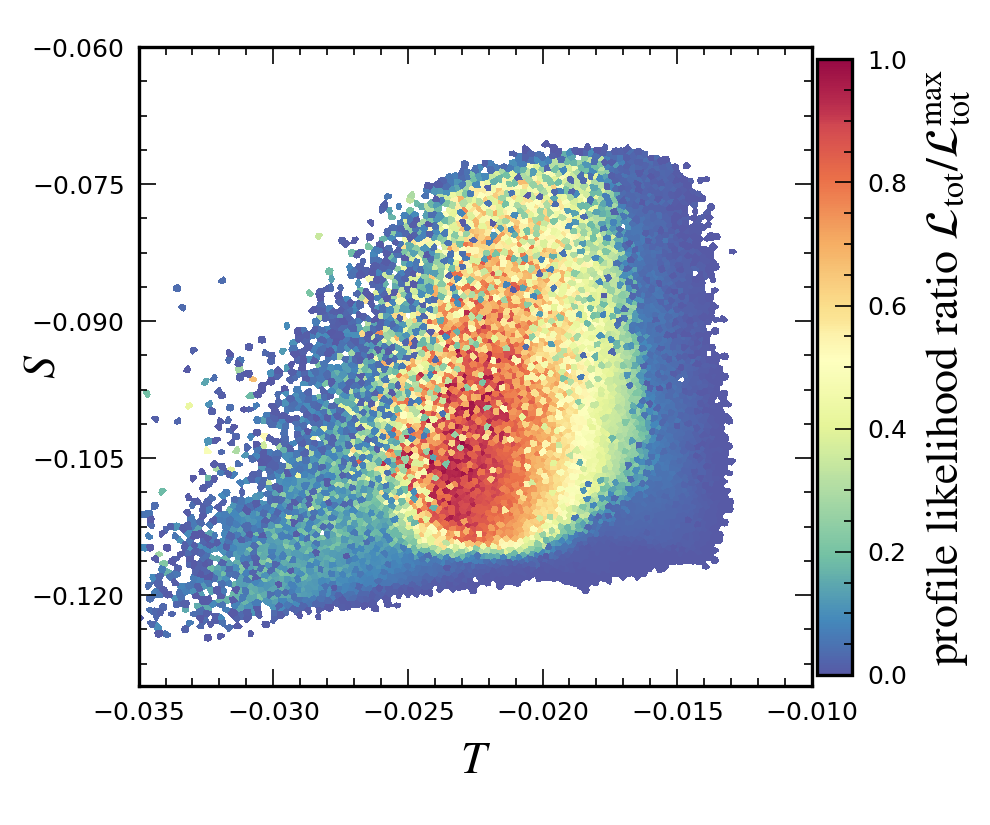}
  \caption{\label{fig:obs1} Similar to Fig.~\ref{fig:excesses}, but shown in the oblique-parameter $(T, S)$ plane and in the branching-ratio plane for the rare $B$-meson decays $B^0_s \to \mu^+ \mu^-$ and $B \to X_s \gamma$. }
\end{figure}
The results for the branching ratios of rare $B$-meson decays and the EW precision observables $T$ and $S$ are shown in Fig.~\ref{fig:obs1}.
These rare $B$-decay processes are highly suppressed in the SM and are dominated mainly by top-quark and $W$-boson loops. In the GM model, new charged scalars such as $H_3^\pm$, $H_5^\pm$, and $H_5^{\pm\pm}$ also participate in these loop diagrams. In our results, the two branching ratios are centered around the experimental central values and exhibit a strong negative correlation. This anti-correlation can be understood at the level of the effective weak Hamiltonian. Both observables are loop-induced flavor-changing neutral-current processes and are therefore sensitive to the charged-scalar sector of the GM model, in particular to the custodial triplet state $H_3^\pm$. The radiative decay $B \to X_s \gamma$ is governed by the electromagnetic dipole operator
\begin{equation}
  \mathcal{H}_{\rm eff} \supset
  -\frac{4G_F}{\sqrt{2}} V_{tb}V_{ts}^*
  C_7^{\rm eff} \mathcal{O}_7, \quad
  \mathcal{O}_7 = \frac{e}{16\pi^2} m_b
  (\bar s \sigma^{\mu\nu} P_R b) F_{\mu\nu}.
\end{equation}
The branching ratio scales approximately as $\mathcal{B}(B \to X_s \gamma) \propto |C_7^{\rm eff}|^2$~\cite{Buras:1998raa, Misiak:2015xwa}.
In contrast, the rare decay $B_s^0 \to \mu^+ \mu^-$ is described by
\begin{equation}
  \mathcal{H}_{\rm eff} \supset -\frac{4G_F}{\sqrt{2}} V_{tb}V_{ts}^*
  \left( C_{10} \mathcal{O}_{10} + C_S \mathcal{O}_S + C_P \mathcal{O}_P \right), \quad
  \mathcal{O}_{10} =
  (\bar s \gamma^\mu P_L b)(\bar\mu\gamma_\mu\gamma_5\mu),
\end{equation}
where $\mathcal{O}_{S,P}$ denote scalar and pseudoscalar operators. The corresponding branching ratio depends predominantly on $|C_{10}|^2$ in the absence of large scalar contributions~\cite{Bobeth:2013uxa,Hermann:2013kca}.
In the GM model, the charged scalars $H_3^\pm$, $H_5^\pm$, and $H_5^{\pm\pm}$ modify both $C_7$ and $C_{10}$ at the one-loop level. However, the loop functions entering $\Delta C_7$ and $\Delta C_{10}$ differ in structure and interfere differently with the SM contributions.  For the parameter region relevant to the 95~GeV interpretation, the charged-scalar contribution tends to shift $C_7$ and $C_{10}$ in opposite effective directions relative to their SM values, leading to the nearly linear anti-correlation observed in Fig.~\ref{fig:obs1}. Turning to the EW oblique measurements, the $T$ parameter is often considered a sensitive probe of custodial symmetry breaking. In the GM model, custodial symmetry is preserved at tree level, so the main contributions to $T$ arise from mass splittings within the triplet and quintuplet states. For all samples, custodial symmetry remains nearly exact, resulting in $T \approx 0$. However, for larger $v_\Delta$ or varying $\alpha$, mixing between the light scalar $H$ and the 125 GeV Higgs can shift the masses of certain scalars, causing mass differences within custodial multiplets and leading to deviations in $T$. The $S$ parameter reflects corrections to the $Z$ boson self-energy and its mixing, originating from new physics states. In the GM model, the $S$ parameter is sensitive to light charged scalars.
When the light scalar $H$ has a larger triplet component, such as in the region $\alpha \sim -\pi/3$, loop effects are enhanced, resulting in more significant deviations in $S$. Conversely, when the 125~GeV Higgs boson is near the alignment limit, such as for $\alpha \sim \pi/3$ and a small triplet vacuum expectation value $v_\Delta$, the scalar sector is dominated by the doublet component, and mixing with the triplet states is suppressed. In this regime, the additional custodial multiplet scalars either become heavier or couple more weakly to the electroweak gauge bosons. As a result, their loop-induced contributions to the gauge-boson self-energies are reduced, leading to smaller corrections to the oblique parameter $S$.
\begin{figure}
  \centering
  \includegraphics[width=0.49\linewidth]{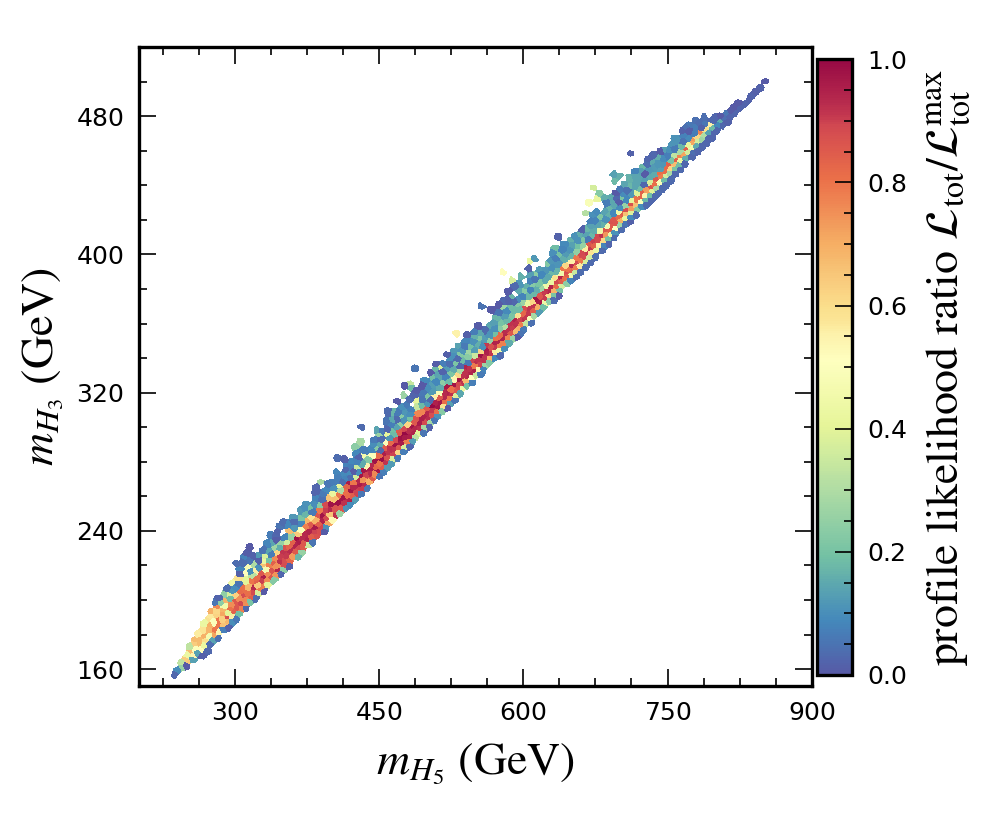}
  \includegraphics[width=0.49\linewidth]{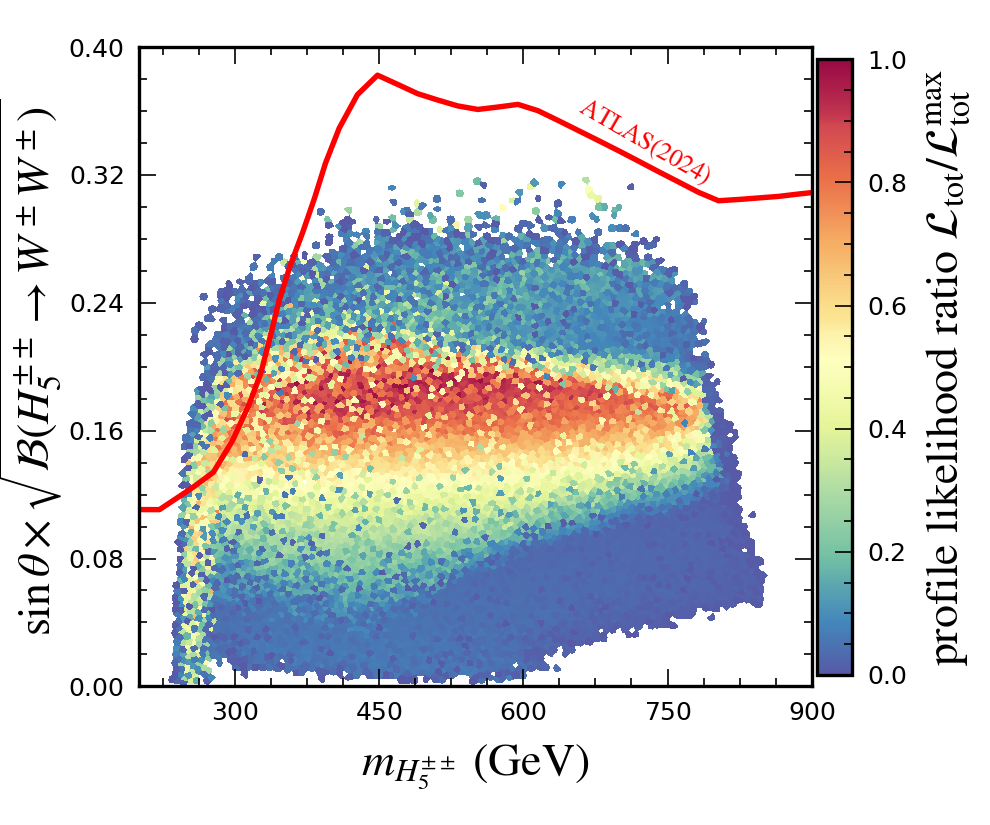}
  \caption{\label{fig:mH3mH5} Similar to Fig.~\ref{fig:excesses}, but showing the $(m_{H_5}, m_{H_3})$ panel and the $m_{H_{5}^{\pm\pm}}$ versus $\sin\theta_H \times \sqrt{\mathcal{B}\left( H_{5}^{\pm\pm} \to W^\pm W^\pm \right)}$ panel. The red line indicates the ATLAS search~\cite{ATLAS:2023sua} limit on doubly charged $H_{5}^{\pm\pm}$ in $W^\pm W^\pm$ final states. }
\end{figure}

\par Finally, Fig.~\ref{fig:mH3mH5} displays the profiled likelihood projected onto the $(M_{H_5},\,M_{H_3})$ plane.
A compressed spectrum is observed between the custodial fiveplet and triplet masses, with $\Delta m(m_{H_5}, m_{H_3}) < \mathcal{O}(100~{\rm GeV})$.
In particular, the robust mass hierarchy $M_{H_5}>M_{H_3}$ implies that the doubly charged scalar $H_5^{\pm\pm}$ tends to decay predominantly into $H_3^\pm W^\pm$, rather than through the conventional $W^\pm W^\pm$ mode.
As shown in Table~\ref{tab:Benchmarks}, the branching ratio $\mathcal{B}(H_5^{\pm\pm}\to W^\pm W^\pm)$ is typically suppressed to below the ten-percent level, and for the benchmark point \texttt{BP2} it can even reach the per-mille level. This feature has important implications for collider phenomenology.

The right panel of Fig.~\ref{fig:mH3mH5} shows the plane of $m_{H_5^{\pm\pm}}$ versus $\sin\theta_H\,\sqrt{\mathcal{B}(H_5^{\pm\pm}\to W^\pm W^\pm)}$, together with the published ATLAS limit~\cite{ATLAS:2023sua} derived for the simplified topology with dominant $H_5^{\pm\pm}\to W^\pm W^\pm$ decays.
We emphasize that this comparison is only indicative. In the cascade-dominated regime relevant to our preferred parameter space, the actual signal topology is instead controlled by $H_5^{\pm\pm}\to H_3^\pm W^\pm$, and the resulting event kinematics and analysis efficiencies need not coincide with those assumed in the published same-sign diboson search. Therefore, the experimental contour in Fig.~\ref{fig:mH3mH5} should not be interpreted as a strict exclusion bound on the full GM scenario in this region.

Nevertheless, the figure still illustrates an important trend: once the branching ratio into $W^\pm W^\pm$ is strongly suppressed, the direct sensitivity of the existing same-sign diboson searches is correspondingly weakened. A dedicated recast of the relevant ATLAS and CMS analyses, explicitly incorporating the cascade topology $H_5^{\pm\pm}\to H_3^\pm W^\pm$, would be required to assess the true collider reach for this part of the parameter space.

% \textcolor{red}{
Additionally, for the Drell–Yan process, which produces $H_5$ pairs via $\gamma$ and $Z$ exchange~\cite{ATLAS:2021jol}, the production cross section depends only on the gauge coupling and the Higgs mass $m_{H_5}$, and is independent of $v_\Delta$ as discussed in the experimental searches. In our work, $m_{H_5}$ ranges from several hundred GeV, where the cross section is relatively small, making it difficult for current experimental searches to exclude the surviving parameter space. The process $H_5^{\pm\pm} \to \ell^{\pm}\ell^{\pm}$ proceeds via a lepton number violation coupling between the triplet $\chi$ and lepton fields, conducted at LEP~\cite{OPAL:2001luy}. However, as pointed out in~\cite{Chiang:2012cn}, this decay channel becomes comparable to the weak gauge boson decay modes only when $v_\Delta \lesssim 10^{-4}~\text{GeV}$, making it negligible in the parameter region of interest for our study.
% }

% Section 5.3: Additional checks
\subsection{Additional checks}

\par
\begin{figure}
  \centering
  \includegraphics[width=0.49\linewidth]{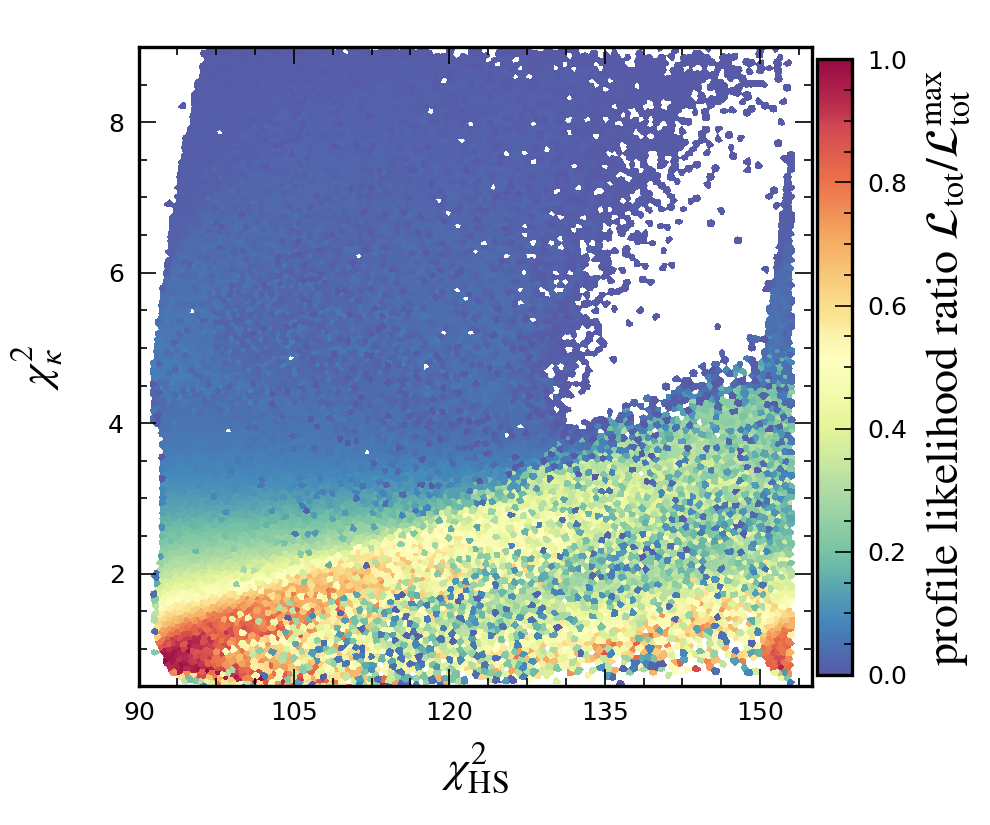}
  \includegraphics[width=0.49\linewidth]{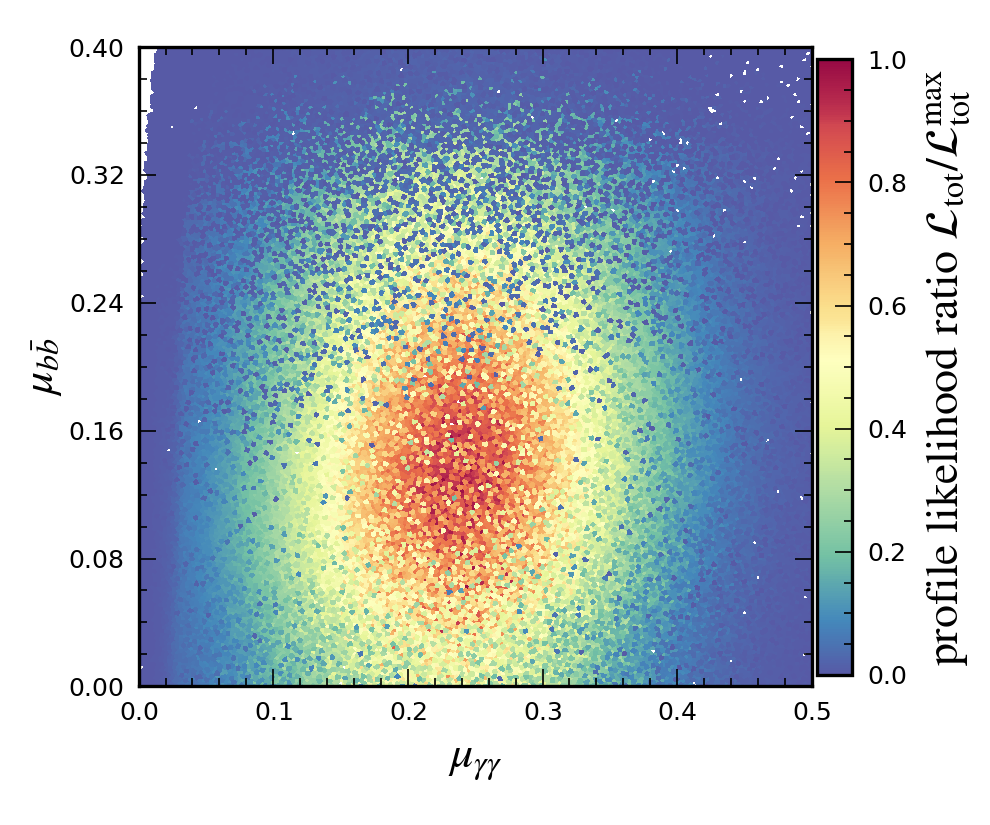}
  \caption{\label{fig:HSbb} Similar to Fig.~\ref{fig:excesses}. Left: relation between $\chi^2_\kappa$ and $\chi^2_{\rm HS}$, where $\chi^2_\kappa$ is derived from the $(\kappa_V,\kappa_F)$ plane analysis and $\chi^2_{\rm HS}$ is computed by the \texttt{HiggsSignals} package. Right: profile-likelihood distribution projected onto the signal strengths in the $\gamma\gamma$ and $b\bar b$ channels.}
\end{figure}

 As discussed in Sec.~\ref{sec-4}, our main fit uses the $(\kappa_V,\kappa_F)$ likelihood rather than the full \texttt{HiggsSignals} output, in order to avoid double counting Higgs signal-strength information. It is nevertheless useful to verify that the preferred region identified in this reduced Higgs-likelihood treatment remains broadly consistent with the more detailed channel-based evaluation provided by \texttt{HiggsSignals}. The left panel of Fig.~\ref{fig:HSbb} shows the relation between $\chi^2_\kappa$ and $\chi^2_{\rm HS}$, with the color scale indicating the total profile-likelihood ratio.

\par One can clearly see that the parameter points with small $\chi^2_\kappa$ also tend to yield comparatively small $\chi^2_{\rm HS}$ values. This result demonstrates that the simplified $(\kappa_V,\kappa_F)$ fit effectively captures the primary Higgs-coupling constraints relevant to our analysis, while \texttt{HiggsSignals} provides a more comprehensive channel-level evaluation. The spread visible in Fig.~\ref{fig:HSbb} indicates that the two approaches are not identical, as anticipated. In particular, the $(\kappa_V,\kappa_F)$ fit is based on a compressed description of the 125~GeV Higgs data and therefore does not retain the full channel-by-channel information, production-mode decomposition, or experimental correlations that are incorporated in \texttt{HiggsSignals}. For this reason, the $\kappa$-plane treatment should be regarded as an effective approximation rather than a complete replacement of the full Higgs likelihood. In practice, it is also somewhat more conservative than the full channel-based \texttt{HiggsSignals} evaluation, since the compressed $(\kappa_V,\kappa_F)$ description retains less experimental information and therefore has weaker discriminating power in excluding parameter points.
Nevertheless, this comparison tests the robustness of our results. The conclusion that the $95~{\rm GeV}$ $\tau\tau$ signal intensity is constrained by $125~{\rm GeV}$ Higgs boson data holds true regardless of whether we use the simplified $\kappa$-plane method or the more detailed channel-based evaluation method.

% \paragraph{Correlation with the LEP $b\bar b$ excess.}

\par The right panel of Fig.~\ref{fig:HSbb} shows a related feature of the surviving parameter space in the $b\bar b$ channel. Although the LEP $b\bar b$ excess is not included as an input constraint in our fit, some viable parameter points remain compatible with the corresponding preferred signal-strength range. In the GM model, the $b\bar b$ and $\tau\tau$ signal strengths are not independent, since both are controlled by the same underlying singlet--doublet mixing structure and share the common factor $|\kappa_F^H|^2 \cdot \Gamma_h^{\rm SM}/\Gamma_H^{\rm tot}$, as discussed in Sec.~\ref{sec-3}. As a result, once the preferred region is selected by the combined $\gamma\gamma$ and $\tau\tau$ fit, the corresponding prediction for $\mu_{b\bar b}$ is automatically restricted to a relatively narrow range.

\par In particular, the best-fit region with $\mu_{\tau\tau}\sim 0.45$ tends to predict a $b\bar b$ signal strength close to the central value of the LEP excess. We do not interpret this as a primary fitting target of the present work, but rather as a correlated outcome of the surviving region selected by the LHC-driven channels. This behavior can be understood qualitatively from the fact that, unlike the $\tau\tau$ excess, the $b\bar b$ signal strength does not encounter the same direct tension with the precision data of the 125~GeV Higgs boson. Consequently, once the Higgs-coupling constraints compress the viable parameter space, the corresponding prediction in the $b\bar b$ channel can still remain close to the LEP preferred range.

\par Our result should not be viewed as being in conflict with earlier studies, but rather as a complementary perspective. Ref.~\cite{Chen:2023bqr} employed a more complete treatment of the Higgs-sector constraints and concluded that the GM model can account for the di-photon excess and remain compatible with the LEP $b\bar b$ excess, but cannot reproduce the large central value of the CMS $\tau\tau$ excess. Ref.~\cite{Ahriche:2023wkj} further showed that a larger $\tau\tau$ rate can arise either in a tiny single-peak region or, more efficiently, in a twin-peak setup involving multiple nearly degenerate states. In contrast, our analysis focuses on a single light $CP$-even state and shows that, within this setup, the $\tau\tau$ signal strength can be pushed to a moderate level of order $0.5$, while larger values are limited by the 125~GeV Higgs data. 

\section{Conclusions}
\label{sec:sum}

In this work we investigated whether a single light $CP$-even custodial singlet in the Georgi--Machacek model can jointly accommodate the LHC hints near $95~\mathrm{GeV}$ in the $\gamma\gamma$ and $\tau\tau$ channels. 
The di-photon excess can be accommodated near its central value, while $\mu_{\tau\tau}\simeq0.5$ lies about $1.4\sigma$ below the CMS value $1.2$ and is compatible only within $2\sigma$.

The one-loop RG-improved positive-definiteness condition reshapes the allowed parameter space relative to the conventional electroweak-scale BFB requirement and yields a net expansion of the phenomenologically viable region, while remaining consistent with perturbative unitarity, electroweak precision data, $B$-physics observables, direct searches and Higgs signal-strength measurements.

The preferred region is characterized by a narrow mixing angle $\alpha\simeq-\pi/3$, a moderate triplet VEV of a few tens of GeV, and a mass hierarchy $m_{H_3}<m_{H_5}$ at the $\mathcal{O}(10^2)~\mathrm{GeV}$ scale. Charged-scalar loops (especially from $H_5^{\pm\pm}$) enhance the di-photon rate, allowing it to approach the experimental central value.

These features make the scenario highly testable. Updated $\gamma\gamma$ and $\tau\tau$ searches at the HL-LHC, together with direct probes of the custodial multiplet states, will cover much of the remaining parameter space. Future lepton colliders, such as CEPC and FCC-ee will provide decisive tests via precision $\kappa_V$ measurements.

\begin{acknowledgments}
We thank the reviewers for their constructive comments. We thank Cheng-Wei Chiang and Martin J. White for helpful discussions. This work was supported by the National Natural Science Foundation of China (Grant Nos. 12447167 and 12275067), by the Joint Fund of Henan Province Science and Technology R$\&$D Program (Grant Nos.225200810030, 225200810092 and 225200810014), by the Startup Research Fund of Henan Academy of Sciences (Grant No. 231820011), by the Basic Research Fund of Henan Academy of Sciences (Grant No. 240620006), by the Graduate Innovation Fund of Henan Academy of Sciences (Grant No. 243320031), by the National Key R$\&$D Program of China (Grant No. 2023YFA1606000) and by the Science and Technology Innovation Leading Talent Support Program of Henan Province  (Grant No. 254000510039).
  PZ is supported by the ARC Discovery Project DP220100007 and by the Centre for the Subatomic Structure of Matter (CSSM).
\end{acknowledgments}

\section*{Data Availability Statement}
All figures presented in this work were generated using the \texttt{Jarvis-PLOT} framework. The complete plotting configuration files used in this study, together with additional result figures not shown in the main text, are publicly available on the webpage:
\begin{center}
  \url{https://pengxuan-zhu-phys.github.io/Jarvis-Docs/jarvisplot/pub/gm95excess/}
\end{center}

\bibliography{reference}
\bibliographystyle{CitationStyle}

\end{document}